\documentclass[lettersize,journal]{IEEEtran}
\usepackage{amsmath,amsfonts}
\usepackage{algorithmic}
\usepackage{algorithm}
\usepackage{array}
\usepackage{amssymb}
\usepackage[caption=false,font=normalsize,labelfont=sf,textfont=sf]{subfig}
\usepackage{textcomp}
\usepackage{stfloats}
\usepackage{url}
\usepackage{verbatim}
\usepackage{graphicx}
\usepackage{cite}
\usepackage{amssymb}
\usepackage{amsthm}
\usepackage{color}
\usepackage{algorithm}
\usepackage{algorithmic}
\usepackage{flushend}
\begin{document}

\title{Delay-Doppler Signal Processing with Zadoff-Chu Sequences}

\author{\IEEEauthorblockN{
Sandesh Rao Mattu\IEEEauthorrefmark{1},
Imran Ali Khan\IEEEauthorrefmark{2},
Venkatesh Khammammetti\IEEEauthorrefmark{1},
Beyza Dabak\IEEEauthorrefmark{1},
Saif Khan Mohammed\IEEEauthorrefmark{2},
Krishna Narayanan\IEEEauthorrefmark{3},
and 
Robert Calderbank\IEEEauthorrefmark{1},~\IEEEmembership{Fellow,~IEEE}}

\IEEEauthorblockA{\IEEEauthorrefmark{1}Department of Electrical and Computer Engineering, Duke University, USA}
\IEEEauthorblockA{\IEEEauthorrefmark{2}Department of Electrical Engineering, Indian Institute of Technology Delhi, India}
\IEEEauthorblockA{\IEEEauthorrefmark{3}Department of Electrical and Computer Engineering, Texas A\&M University, USA}
\thanks{The Duke and Texas A\&M teams are supported in part by the National Science Foundation under grants 2148212 and 2148354 respectively and both teams are supported in part by funds from federal agency and industry partners as specified in the Resilient \& Intelligent NextG Systems (RINGS) program. The partnership between the Duke team and the IIT Delhi team is supported by the US National Science Foundation under grant 2342690 and by the Department of Science and Technology, Govt. of India under grant TPN-96226. The Duke team is supported in part by the Air Force Office of Scientific Research under grants FA 8750-20-2-0504 and FA 9550-23-1-0249. S. K. Mohammed is also associated with Bharti School of Telecom. Technology and Management (BSTTM), IIT Delhi. The work of S. K. Mohammed is
supported in part by the Jai Gupta Chair at IIT Delhi.

This work may be submitted to the IEEE for possible publication. Copyright may be transferred without notice, after which this version may no longer be accessible.
}
}



\maketitle

\begin{abstract}
Much of the engineering behind current wireless systems has focused on designing an efficient and high-throughput downlink to support human-centric communication such as video streaming and internet browsing. This paper looks ahead to design of the uplink, anticipating the emergence of machine-type communication (MTC) and the confluence of sensing, communication, and distributed learning. We demonstrate that grant-free multiple access is possible even in the presence of highly time-varying channels. Our approach provides a pathway to standards adoption, since it is built on enhancing the 2-step random access procedure which is already part of the 5GNR standard. This 2-step procedure uses Zadoff-Chu (ZC) sequences as preambles that point to radio resources which are then used to upload data. We also use ZC sequences as preambles / pilots, but we process signals in the Delay-Doppler (DD) domain rather than the time-domain. We demonstrate that it is possible to detect multiple preambles in the presence of mobility and delay spread using a receiver with no knowledge of the channel other than the worst case delay and Doppler spreads. Our approach depends on the mathematical properties of ZC sequences in the DD domain. We derive a closed form expression for ZC pilots in the DD domain, we characterize the possible self-ambiguity functions, and we determine the magnitude of the possible cross-ambiguity functions. These mathematical properties enable detection of multiple pilots through solution of a compressed sensing problem. The columns of the compressed sensing matrix are the translates of individual ZC pilots in delay and Doppler. We show that columns in the design matrix satisfy a coherence property that makes it possible to detect multiple preambles in a single Zak-OTFS subframe using One-Step Thresholding (OST), which is an algorithm with low complexity.

We transmit data using Zak-OTFS modulation, where the carrier waveforms are pulses in the DD domain regularly spaced on a period grid. We describe how mathematical properties of ZC spread pilots make it possible to integrate sensing and communication within a single Zak-OTFS subframe through the combination of a ZC spread pilot used to sense the I/O relation and point pulsones for data transmission. We emphasize that there is no division of radio resources between sensing the I/O relation on the period lattice and data transmission, that the two functions coexist in the same Zak-OTFS subframe. 
\end{abstract}

\begin{IEEEkeywords}
Wireless networks, Unsourced random access, Compressed sensing, Zadoff-Chu sequences, OTFS modulation
\end{IEEEkeywords}

\section{Introduction}
\label{sec:introduction}
\IEEEPARstart{C}{urrent} wireless infrastructure revolves around a high-throughput downlink \cite{liu2001opportunistic,qin2006distributed,eryilmaz2007fair,choi2007opportunistic,hou2009theory,gopalan2012wireless} optimized to support human-centric applications such as voice telephony, internet browsing, navigation, and music / video streaming.

Emerging applications such as automated driving, augmented reality (AR) and teleconferencing require live streaming from mobile devices, and as users become sources of content, engineering focus is shifting to the uplink. 
As federated learning and distributed machine learning applications become more prevalent, there will be a growing need to support Mobile Massive Machine Type Communications (3MTC). 
This paper focuses on the challenge of providing low latency and ubiquitous connectivity for the combination of short packets, unpredictable traffic patterns, and high Doppler spreads that is characteristic of 3MTC.

At the medium access control (MAC) layer, unpredictable traffic patterns render the current paradigm of scheduling users on reserved uplink resources (grants) very inefficient. 
Random access at the MAC layer is an alternative to scheduling users, and Release 15, 3GPP introduced the 2-step RACH as a mechanism to enable grant-free (random) initial access. 
Since then, there has been considerable interest in supporting multiple traffic types including massive multiple access using the 2-step RACH procedure \cite{2-step-workitem,peralta2021two,agostini2024evolution}.

However, the throughput and delay of naive random-access techniques scale very poorly with the number of users. Indeed, Polyanskiy \cite{polyanskiy2017perspective} showed that there is a very large gap between the performance of naive slotted ALOHA and what is information-theoretically achievable through random coding. Since then, unsourced random access has emerged as an effective solution for massive connectivity ~\cite{zucchetto2017uncoordinated,liu2018sparse,munari2020grant,chen2020massive,liva2024unsourced}. 

Several schemes such as Coded slotted ALOHA \cite{liva2011graph,vem2019user}, sparse IDMA \cite{pradhan2022sparse}, coded compressed sensing \cite{amalladinne2020unsourced,fengler2021sparcs,calderbank2018chirrup}, and polar coding based unsourced random access \cite{pradhan2020polar,marshakov2019polar,ahmadi2021random} have been developed and they have been shown to perform close to the information-theoretic benchmark.
Recently, the authors in the white paper \cite{agostini2024evolution} present a comprehensive analysis of how unsourced random access can enhance the 2-step RACH procedure for grant-free access. This paper explores how unsourced random access based on the 2-step RACH might form a blueprint for next generation multiple access. The first of the two steps is preamble detection, and here we demonstrate simultaneous detection of multiple preambles in the presence of mobility and delay spread. The second step is uploading data, and here we demonstrate integration of channel sensing and data transmission, again in the presence of mobility and delay spread.

When analyzing the potential of random coding Polyanskiy \cite{polyanskiy2017perspective} made the simplifying assumption of a Gaussian random-access channel. Subsequently, sophisticated unsourced random access methods have been designed and evaluated primarily for channels without mobility. 
This is a major limitation given the potential overhead in estimating rapidly time-varying channels. This limitation becomes more serious given the interest in using the FR3 band (6 GHz -- 24 GHz) for supporting 6G cellular communications, and the interest in non-terrestrial networks (NTNs) where the receiver is typically moving at very high velocities. High Doppler spreads are characteristic of both these developments.

6G propagation environments are changing the balance between time-frequency methods focused on OFDM signal processing and delay-Doppler methods. OFDM is configured to prevent inter-carrier-interference (ICI) whereas Zak-OTFS is configured to embrace ICI. In OFDM, once the I/O relation is known, equalization is relatively simple at least when there is no ICI. However, acquisition of the I/O relation is non-trivial and model dependent. In contrast, equalization is more involved in Zak-OTFS due to inter-symbol-interference (ISI), however acquisition of the I/O relation is simple and model free. Acquisition becomes more critical when Doppler spreads measured in KHz make it more and more difficult to estimate channels. The numerical simulations in Sections V and VI are performed for the most challenging situation, which is the combination of unresolvable paths and high channel spreads. In this scenario, multicarrier approaches that are based on channel estimation break down, and Zak-OTFS has been shown to deliver superior performance (\cite{OTFS2Paper2}, Fig. 17).

Section \ref{sec:domains} reviews the parametric family of Zak-OTFS carrier (pulsone) waveforms that can be matched to the delay and Doppler spreads of different propagation environments. We describe how the (point) pulsone signal in the time domain realizes a quasi-periodic localized function on the DD domain. The characteristic structure of a pulsone is a train of pulses modulated by a tone, a signal with unattractive peak-to-average power ratio (PAPR). Section \ref{sec:system_model} describes how the I/O relation of the sampled communication system can be read off from the response to the point pilot pulsone used to probe the effective channel, which incorporates transmit and receive filtering. We refer the reader to \cite{OTFS2Paper2} for a review of system performance in the crystalline regime where the delay period of the pulsone is greater than the delay spread of the channel, and the Doppler period of the pulsone is greater than the Doppler spread of the channel. 

When channel sensing and data transmission take place in separate subframes, the point pulsone used to sense the I/O relation does not interfere with the point pulsones used to transmit data. Section \ref{sec:pilot_design} reviews the method of filtering in the discrete DD domain, introduced in \cite{ISAC}. We describe how a chirp filter in the discrete DD domain is applied to a point pulsone to produce a spread waveform with desirable characteristics. One desirable characteristic is low PAPR, about 6 dB for the exemplar spread pulsone, compared with about 15 dB for the point pulsone. A second desirable characteristic is the ability to read off the I/O relation of the sampled communication system provided a second crystallization condition is satisfied. We then present Zadoff-Chu (ZC) spread pilots as an alternative to Zak-OTFS spread pilots, with the same low PAPR as Zak-OTFS spread pilots. Our approach to grant free multiple access depends on mathematical properties of ZC sequences. In Section \ref{sec:pilot_design} we derive a closed form expression for Zadoff-Chu (ZC) spread pilots in the DD domain, we characterize the support of the possible self-ambiguity functions, and we determine the magnitude of the possible cross-ambiguity functions.

In Section \ref{sec:datatransmission} we describe how the mathematical properties of ZC spread pilots make it possible to integrate sensing and communication within a single Zak-OTFS subframe through the combination of a ZC spread pilot used to sense the I/O relation and point pulsones for data transmission. An alternative approach is to sense the I/O relation with a point pulsone surrounded by a guard band where no data is transmitted. This approach divides the resources between sensing and data transmission, thereby limiting the effective throughput. In our approach, there is no division of radio resources between sensing the I/O relation on the period lattice and data transmission, that the two functions coexist in the same Zak-OTFS subframe. We refer the reader to \cite{ISAC} for a parallel development using Zak-OTFS spread pilots. The numerical simulations presented in Section \ref{sec:datatransmission} show that there is essentially no difference between sensing the effective channel with Zak-OTFS spread pilots and sensing the effective channel with ZC spread pilots.

There is still a significant performance gap between our approach and sending the I/O relation and transmitting data in separate Zak-OTFS subframes. We conclude Section \ref{sec:datatransmission} by showing that turbo signal processing can close this performance gap. We refer the reader to \cite{turbo_paper} for a parallel development using Zak-OTFS spread pilots.

In Section \ref{sec:preamble_detection} we describe how the mathematical properties of ZC spread pilots enable detection of multiple pilots through solution of a compressed sensing problem. The columns of the compressed sensing/design matrix are the translates of individual pilots in delay and Doppler on the period lattice. The receiver has no knowledge of the physical channel other than the worst case delay and Doppler spreads. The design matrix satisfies the coherence property introduced in \cite{BCJ}, making it possible to use One-Step Thresholding (OST) to detect multiple preambles.

We emphasize that the Veh-A channel; model used in our numerical simulations introduces fractional delay and Doppler representative of real propagation environments. When we detect multiple preambles the receiver has no information about individual path delays. We are solving a compressed sensing problem where there is a mismatch between the assumed and the actual bases for sparsity. We refer the reader to \cite{cspc} for a discussion of basis mismatch.

Section \ref{sec:conclusions} presents conclusions and directions for future work.

\section{Time, Frequency, and Delay-Doppler domains}
\label{sec:domains}
It is common knowledge that a time-domain (TD) pulse is an ideal waveform for pure delay channels as it allows to separate reflections according to their range, and, similarly, a frequency domain (FD) pulse is an ideal waveform for pure Doppler channels as it allows to separate reflections according to their velocity. We will demonstrate that a pulse in the DD domain is an ideal waveform for doubly spread channels consisting of reflections of various ranges and velocities. 

\begin{figure*}[ht]
    \centering
    \includegraphics[clip=true, trim=0in 0in 0in 0.4in, width=0.8\textwidth]{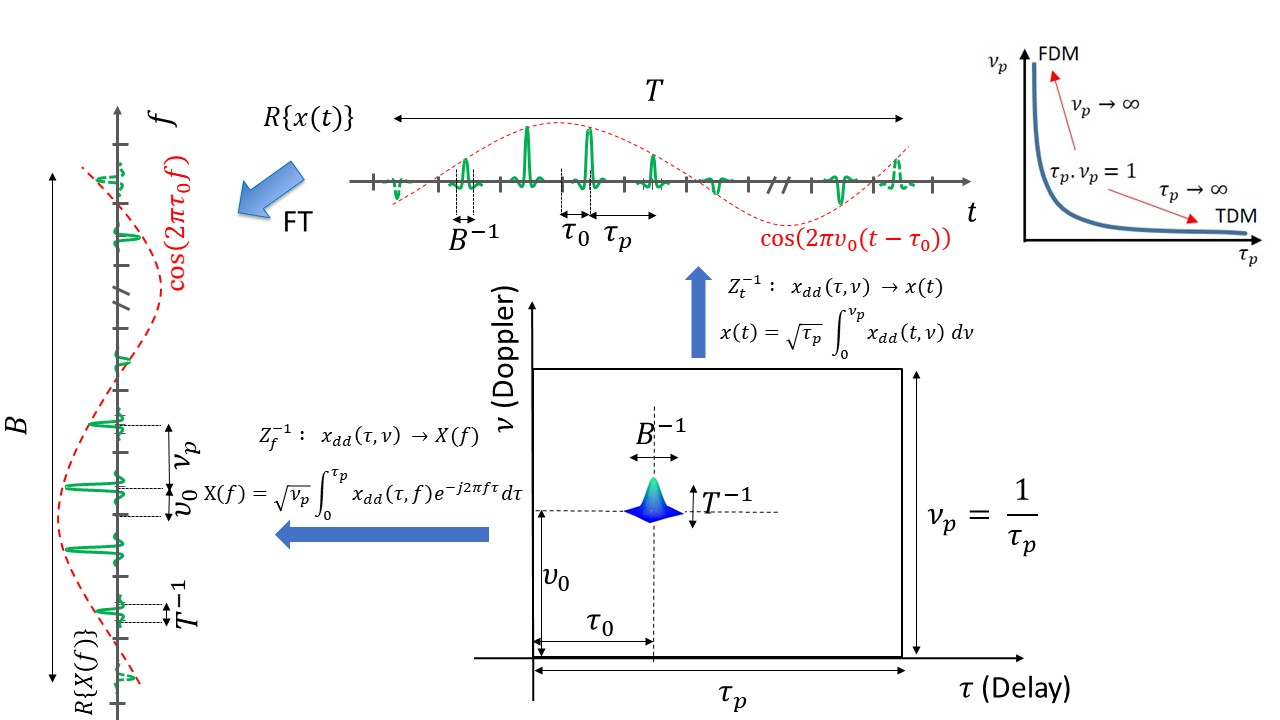}
    \caption{ A DD domain pulse and its TD/FD realizations referred to as TD/FD pulsone. The TD pulsone comprises of a finite duration pulse train modulated by a TD tone. The FD pulsone comprises of a finite bandwidth pulse train modulated by a FD tone. The location of the pulses in the TD/FD pulse train and the frequency of the modulated TD/FD tone is determined by the location of the DD domain pulse $(\tau_0, \nu_0)$. The time duration ($T$) and bandwidth ($B$) of a pulsone are inversely proportional to the characteristic width of the DD domain pulse along the Doppler axis and the delay axis, respectively. The number of non-overlapping DD pulses, each spread over an area $B^{-1}T^{-1}$, inside the fundamental period $\mathcal{D}_0$ (which has unit area) is equal to the time-bandwidth product $BT$ and the corresponding pulsones are orthogonal to one another, rendering OTFS an orthogonal modulation that achieves the Nyquist rate. As the Doppler period $\nu_p\to\infty$, the FD pulsone approaches a single FD pulse which is the FDM carrier. Similarly, as the delay period $\tau_p\to\infty$, the TD pulsone approaches a single TD pulse which is the TDM carrier. Setting $\tau_p\nu_p = 1$, we see that OTFS is a family of modulations parameterized by $\tau_p$ that interpolates between TDM and FDM.}
    \label{fig4paper1} 
\end{figure*}

A pulse in the DD domain is a quasi-periodic localized function, defined by a delay period $\tau_p$ and a Doppler period $\nu_p$.
In the \emph{period lattice} $\Lambda_p = \{ (n \tau_p, m \nu_p) \, | \, n,m \in {\mathbb Z} \}$, there is only one pulse within the \emph{fundamental region} ${\mathcal D}_0 = \{ (\tau, \nu) \, | \, 0 \leq \tau < \tau_p, 0 \leq \nu < \nu_p \},$ and there are infinitely many replicas along the delay and Doppler axes given by
\begin{eqnarray}
    x_{_{\mbox{\scriptsize{dd}}}}(\tau + n \tau_p,\nu + m \nu_p)  & = &   e^{j 2 \pi n \nu \tau_p} \, x_{_{\mbox{\scriptsize{dd}}}}(\tau,\nu),
\end{eqnarray} for all $n,m \in {\mathbb Z}$. Only
quasi-periodic DD domain functions can have a TD representation. When viewed in the TD, this function is realized as a pulse train modulated by a tone (see Fig.~\ref{fig4paper1}), hence the name \textit{pulsone}. The DD domain pulse is the orthogonal time frequency space (Zak-OTFS) waveform, introduced in~\cite{Hadani2017}, and widely studied thereafter (see~\cite{bestreads} and the references therein).

We have shown in \cite{OTFS2Paper1} that the parametric family of pulsone waveforms includes time-division multiplexing (TDM) and frequency-division multiplexing (FDM) waveforms as special cases. TDM waveforms, perfectly matched to delay-only propagation environments, are pulsones with indefinitely large delay period and vanishingly small Doppler period. FDM waveforms, perfectly matched to Doppler-only propagation environments, are pulsones with indefinitely large Doppler period and vanishingly small delay period.

\section{Zak-OTFS System Model}
\label{sec:system_model}
\begin{figure*}
\centering
\includegraphics[width=16.5cm, height=4.5cm]{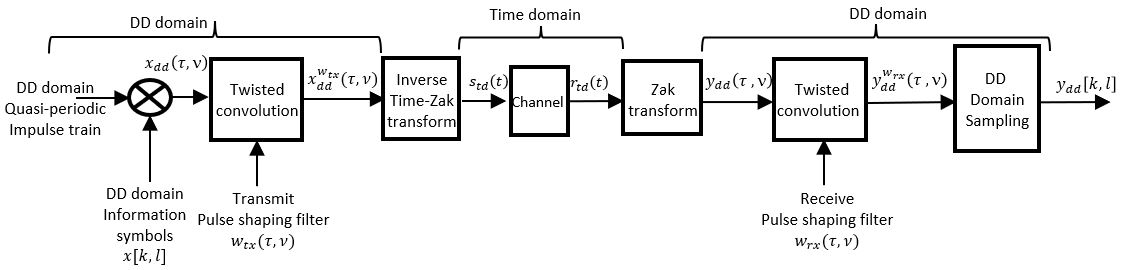}
\caption{Zak-OTFS transceiver processing.} 
\label{figzakotfspaper2}
\end{figure*}
Zak-OTFS transceiver processing is illustrated in Fig.~\ref{figzakotfspaper2} (see Section II of \cite{OTFS2Paper2} also).
We transmit $MN$ symbols $x[k_0,l_0]$ in each subframe, $k_0=0,1,\cdots, M-1$, $l_0=0,1,\cdots, N-1$. The discrete DD domain pulse $x_{\mbox{\scriptsize{dd}}}^{(k_0, l_0)}[k,l]$ carries the information symbol $x[k_0, l_0]$, i.e.
 \begin{align}
 \label{eqn3p4}
     x_{\mbox{\scriptsize{dd}}}^{(k_0, l_0)}[k,l]& = \nonumber \\
     &\hspace{-13mm}\sum\limits_{n,m \in {\mathbb Z}} \hspace{-2mm} {\Big (} e^{j 2 \pi \frac{ n  l_0}{N}} \, x[k_0  , l_0] \, \delta[k - k_0 - nM] \delta[l - l_0 - mN]{\Big )},
 \end{align}
 for all $k,l \in {\mathbb Z}$. From (\ref{eqn3p4}) it is clear that this pulse carrying the $(k_0, l_0)$-th information symbol consists of infinitely many Dirac-delta impulses at discrete DD locations $(k_0 + nM, l_0 + mN)$, $n,m \in {\mathbb Z}$. Note that $x_{\mbox{\scriptsize{dd}}}^{(k_0, l_0)}[k,l]$ is a \emph{quasi-periodic} function with period $M$ along the delay axis and period $N$ along the Doppler axis so that for any $n,m \in {\mathbb Z}$, $x_{\mbox{\scriptsize{dd}}}^{(k_0, l_0)}[k,l]$ satisfies 
 \begin{eqnarray}
     \label{qpeqn}
     x_{\mbox{\scriptsize{dd}}}^{(k_0, l_0)}[k + nM,l + mN] & = & e^{j 2 \pi \frac{n l_0}{N}} \, x_{\mbox{\scriptsize{dd}}}^{(k_0, l_0)}[k,l].
 \end{eqnarray}The discrete DD domain pulses corresponding to all $MN$ information symbols are superimposed resulting in the quasi-periodic discrete DD domain signal
 \begin{eqnarray}
     x_{\mbox{\scriptsize{dd}}}[k,l] & = & \sum\limits_{k_0=0}^{M-1} \sum\limits_{l_0 = 0}^{N-1} x_{\mbox{\scriptsize{dd}}}^{(k_0, l_0)}[k,l].
 \end{eqnarray}
 $x_{\mbox{\scriptsize{dd}}}[k,l]$ is supported on the \emph{information lattice} $\Lambda_{dd} = \{\left( \frac{k \tau_p}{M} , \frac{l \nu_p}{N} \right) \, | \, k, l \in {\mathbb Z} \}$. We then lift the discrete signal $x_{\mbox{\scriptsize{dd}}}[k,l]$ to the continuous DD domain signal 
 \begin{eqnarray}
 \label{eqn2}
     x_{\mbox{\scriptsize{dd}}}(\tau, \nu) & = & \sum\limits_{k,l \in {\mathbb Z}} x_{\mbox{\scriptsize{dd}}}[k,l] \, \delta\left(\tau -  \frac{k \tau_p}{M} \right) \, \delta\left(\nu -  \frac{l \nu_p}{N} \right).
 \end{eqnarray}Note that, for any $n,m \in {\mathbb Z}$, \,\
     $x_{\mbox{\scriptsize{dd}}}(\tau + n \tau_p, \nu + m \nu_p) = e^{j 2 \pi n \nu \tau_p} \, x_{\mbox{\scriptsize{dd}}}(\tau, \nu)$,
 so that $x_{\mbox{\scriptsize{dd}}}(\tau, \nu)$ is periodic with period $\nu_p$ along the Doppler axis and quasi-periodic with period $\tau_p$ along the delay axis. 

 We use a pulse shaping filter $w_{tx}(\tau, \nu)$
 to limit the TD Zak-OTFS subframe to time duration $T = N \tau_p$ and bandwidth $B = M \nu_p$.
  The DD domain transmit signal $x_{\mbox{\scriptsize{dd}}}^{w_{tx}}(\tau, \nu)$ is given by the twisted convolution\footnote{\footnotesize{$a(\tau, \nu) *_{\sigma} b(\tau, \nu) = \iint a(\tau', \nu') \, b(\tau - \tau', \nu - \nu') \, e^{j 2 \pi \nu'(\tau - \tau')} \, d\tau' \, d\nu' $, for any two DD functions, $a(\tau, \nu), b(\tau, \nu)$.}} of the transmit pulse shaping filter $w_{tx}(\tau, \nu)$ with $x_{\mbox{\scriptsize{dd}}}(\tau, \nu)$.
 \begin{eqnarray}
 \label{eqn61094}
     x_{\mbox{\scriptsize{dd}}}^{w_{tx}}(\tau, \nu) & = & w_{tx}(\tau, \nu) \, *_{\sigma} \, x_{\mbox{\scriptsize{dd}}}(\tau, \nu),
 \end{eqnarray}where $*_{\sigma}$ denotes the twisted convolution operator \cite{OTFS2Paper1, OTFS2Paper2}.
 The TD realization of $x_{\mbox{\scriptsize{dd}}}^{w_{tx}}(\tau, \nu)$
 gives the transmitted TD signal which is given by
 \begin{eqnarray}
 \label{eqn8p4}
 s_{\mbox{\scriptsize{td}}}(t) & = & {\mathcal Z}_t^{-1}\left( x_{\mbox{\scriptsize{dd}}}^{w_{tx}}(\tau, \nu) \right),
 \end{eqnarray}where ${\mathcal Z}_t^{-1}$ denotes the inverse Zak transform (see Eqn.~$(7)$ in \cite{OTFS2Paper1} for more details).
The received TD signal is given by
\begin{eqnarray}
\label{eqn712446}
    r_{\mbox{\scriptsize{td}}}(t) & \hspace{-3mm} = & \hspace{-3mm} \iint h_{\mbox{\scriptsize{phy}}}(\tau, \nu) \, s_{\mbox{\scriptsize{td}}}(t - \tau) \, e^{j 2 \pi \nu (t - \tau)} \, d\tau \, d\nu \, + \, n_{\mbox{\scriptsize{td}}}(t),
\end{eqnarray}where $h_{\mbox{\scriptsize{phy}}}(\tau, \nu)$ is the delay-Doppler spreading function of the physical channel and $n_{\mbox{\scriptsize{td}}}(t)$ is AWGN.
At the receiver, we pass from the TD to the DD domain by applying the Zak transform ${\mathcal Z}_t$ to the received TD signal $r_{\mbox{\scriptsize{td}}}(t)$, and we obtain
\begin{eqnarray}
\label{eqn712448}
    y_{\mbox{\scriptsize{dd}}}(\tau, \nu) & = & {\mathcal Z}_t\left( r_{\mbox{\scriptsize{td}}}(t) \right).
\end{eqnarray}Substituting (\ref{eqn712446}) into (\ref{eqn712448}) it follows that \cite{OTFS2Paper1, OTFS2Paper2}
\begin{eqnarray}
    y_{\mbox{\scriptsize{dd}}}(\tau, \nu) & = &  h_{\mbox{\scriptsize{phy}}}(\tau, \nu) \, *_{\sigma} \, x_{\mbox{\scriptsize{dd}}}^{w_{tx}}(\tau, \nu) \, + \, n_{\mbox{\scriptsize{dd}}}(\tau, \nu),
\end{eqnarray}where $n_{\mbox{\scriptsize{dd}}}(\tau, \nu)$ is the DD representation of the AWGN. Note that, in the DD domain the channel acts on the input $x_{\mbox{\scriptsize{dd}}}^{w_{tx}}(\tau, \nu)$ through twisted convolution with $h_{\mbox{\scriptsize{phy}}}(\tau, \nu)$. This is similar to how in linear time invariant (LTI) channels (i.e., delay-only channels), the channel acts on a TD input through linear convolution with the TD channel impulse response.
Twisted convolution is the generalization of linear convolution for doubly-spread channels. 

Next, we apply a matched filter $w_{rx}(\tau, \nu)$ which acts by twisted convolution on $y_{\mbox{\scriptsize{dd}}}(\tau, \nu)$ to give
\begin{eqnarray}
    y_{\mbox{\scriptsize{dd}}}^{w_{rx}}(\tau, \nu) & = & w_{rx}(\tau, \nu) *_{\sigma} y_{\mbox{\scriptsize{dd}}}(\tau, \nu).
\end{eqnarray}This filtered signal is then sampled on the information lattice $\Lambda_{\mbox{\scriptsize{dd}}}$ resulting in the quasi-periodic discrete DD domain signal $y_{\mbox{\scriptsize{dd}}}[k,l]$.
This discrete DD output signal is related to the input discrete DD signal
$x_{\mbox{\scriptsize{dd}}}[k,l]$ through the input-output (I/O) relation \cite{OTFS2Paper1, OTFS2Paper2}
\begin{eqnarray}
\label{eqnio1}
    y_{\mbox{\scriptsize{dd}}}[k,l] & = & h_{\mbox{\scriptsize{eff}}}[k,l] \, *_{\sigma} \, x_{\mbox{\scriptsize{dd}}}[k,l] \, + \, n_{\mbox{\scriptsize{dd}}}[k,l],
\end{eqnarray}where\footnote{\footnotesize{For any two discrete DD functions $a[k,l]$ and $b[k,l]$, the discrete twisted convolution $a[k,l] *_{\sigma} b[k,l] = \sum\limits_{k',l' \in {\mathbb Z}} a[k', l'] \, b[k - k', l- l'] \, e^{j 2 \pi l' \frac{(k - k')}{MN}}$.}} $h_{\mbox{\scriptsize{eff}}}[k,l]$ is the effective DD domain channel filter and $n_{\mbox{\scriptsize{dd}}}[k,l]$ are the DD domain noise samples. Note that $h_{\mbox{\scriptsize{eff}}}k,l]$ is simply
\begin{eqnarray}
    h_{\mbox{\scriptsize{eff}}}(\tau,\nu) = w_{rx}(\tau, \nu) *_{\sigma} h_{\mbox{\scriptsize{phy}}}(\tau, \nu) *_{\sigma} w_{tx}(\tau, \nu) 
    \label{h_eff}
\end{eqnarray}sampled on the information lattice $\Lambda_{\mbox{\scriptsize{dd}}}$, i.e.
\begin{eqnarray}
    h_{\mbox{\scriptsize{eff}}}[k,l] & \triangleq & h_{\mbox{\scriptsize{eff}}}{\Big (}  \tau = \frac{k \tau_p}{M} , \nu = \frac{l \nu_p}{N} {\Big )}.
\end{eqnarray}From the I/O relation in (\ref{eqnio1}) it is clear that in order to detect the DD domain information symbols from $y_{\mbox{\scriptsize{dd}}}[k,l]$, it suffices to have knowledge of
$h_{\mbox{\scriptsize{eff}}}[k,l]$ only. The receiver does not need to acquire $h_{\mbox{\scriptsize{phy}}}(\tau, \nu)$. Instead, it acquires $h_{\mbox{\scriptsize{eff}}}[k,l]$ directly from the channel response to pilots in the discrete DD domain. This makes the Zak-OTFS I/O relation applicable to any model of the underlying physical channel and is therefore \emph{model-free}. Next, we consider acquisition of $h_{\mbox{\scriptsize{eff}}}[k,l]$.

\section{Pilot Design}
\label{sec:pilot_design}
We transmit a pilot signal $x_{\mbox{\scriptsize{p,dd}}}[k,l]$ within a single Zak-OTFS subframe. This signal is quasi-periodic, hence is completely specified by the values it takes within the \emph{fundamental region} ${\mathcal D} = \{ (k,l) \, | \, k=0,1,\cdots, M-1, l =0,1,\cdots, N-1\}$.

\textbf{Zak-OTFS Point Pilots:} Here the pilot signal $x_{\mbox{\scriptsize{p,dd}}}[k,l]$ is determined by a unit energy Dirac-delta impulse at the pilot location $(k_p, l_p) \in {\mathcal D}$ and repeats along the delay and Doppler axis by integer multiples of the delay and Doppler period respectively. It is given by
\begin{align}
\label{eqnexprxpdd1}
    x_{\mbox{\scriptsize{p,dd}}}[k,l] =  \sum\limits_{n,m \in {\mathbb Z}}  e^{j 2 \pi n \frac{l_p}{N}} \, \delta[k - k_p - nM] \, \delta[l - l_p - mN].
\end{align}
From (\ref{eqnio1}) and (\ref{eqnexprxpdd1}), the received pilot is given by
\begin{eqnarray}
\label{eqn83ggd70}
    h_{\mbox{\scriptsize{eff}}}[k,l] *_{\sigma} \left( \sqrt{E_p} \, x_{\mbox{\scriptsize{p,dd}}}[k,l] \right) &  \hspace{-3mm} =  &  \hspace{-3mm} \sqrt{E_p} \hspace{-1mm} \sum\limits_{n,m \in {\mathbb Z}} h_{n,m}[k,l].
\end{eqnarray}where $\sqrt{E_p}$ is the pilot energy. The $(n,m)$-th term $h_{n,m}[k,l]$ is the channel response to the Dirac-delta impulse of the quasi-periodic pilot signal $x_{\mbox{\scriptsize{p,dd}}}[k,l]$ located at $(k_p + nM, l_p + mN)$, and is given by
\begin{align}
\label{eqnhnm}
 h_{\mbox{\scriptsize{eff}}}[k,l] &*_{\sigma} \, \left( e^{j 2 \pi n \frac{l_p}{N}} \delta[k - k_p -nM] \, \delta[l - l_p - mN] \right) \nonumber \\
    & = {\Big (} h_{\mbox{\scriptsize{eff}}}[k -k_p - nM,l - l_p - mN]  \, e^{j 2 \pi \frac{n l_p}{N}} \times \nonumber \\
    & \hspace{6mm} e^{j 2 \pi \frac{(l - l_p -mN) (k_p + nM) }{MN}} {\Big )}.
\end{align}
The support ${\mathcal S}_{n,m}$ of $h_{n,m}[k,l]$ is ${\mathcal S} + (k_p + nM, l_p + mN)$. 
The \emph{crystallization condition} is ${\mathcal S}_{n,m} \cap {\mathcal S}_{n',m'} = \phi$ for $(n,m) \ne (n',m')$, and when it is satisfied, there is no DD domain aliasing. We have emphasized in \cite{OTFS2Paper2}
that the crystallization condition is satisfied when\footnote{\footnotesize{For any real number $x \in {\mathbb R}$, $\lceil x \rceil$ is the smallest integer greater than or equal to $x$.}}
\begin{eqnarray}
\label{cryscnd}
    k_{max} \triangleq \left\lceil \frac{M \tau_{max}}{\tau_p} \right\rceil  <  M, \,\ l_{max} \triangleq \left\lceil \frac{2 N \nu_{max}}{\nu_p} \right\rceil  <  N.
\end{eqnarray}
Here $\tau_{max} > 0$ and $\nu_{max} > 0$ are respectively the maximum possible delay and Doppler shift induced by any physical channel path.
The first condition in (\ref{cryscnd}) is that the channel delay spread $\tau_{max}$ is less than the delay period $\tau_p$, and the second condition is that the channel Doppler spread $2 \nu_{max}$ is less than the Doppler period $\nu_p$. We have shown that non-predictability and fading result from aliasing in the delay-Doppler domain.
(See \cite{OTFS2Paper2}, Section II for more details.) \\
When the crystallization condition holds
\begin{eqnarray}
\hspace{-25mm}  h_{0,0}[k,l] & = & h_{\mbox{\scriptsize{eff}}}[k -k_p,l - l_p]  \, e^{j 2 \pi \frac{k_p(l - l_p)}{MN}}
\end{eqnarray}for $(k,l) \in (k_p, l_p) + {\mathcal S}$ and therefore
\begin{eqnarray}
h_{\mbox{\scriptsize{eff}}}[k,l] & = & h_{0,0}[k+k_p,l+l_p] \, e^{-j 2 \pi \frac{k_p l}{MN}}
\end{eqnarray} for $(k,l) \in {\mathcal S}$. For $(k,l) \in {\mathcal S} + (k_p, l_p)$, the received pilot response (AWGN-free) is simply $h_{0,0}[k,l]$ since the support sets of $h_{n,m}[k,l]$, $n,m \in {\mathbb Z}$ do not overlap when the crystallization condition is satisfied. Hence, the taps of the effective channel filter can simply be read off from the received pilot response within ${\mathcal S} + (k_p, l_p)$. As a result the Zak-OTFS I/O relation in (\ref{eqnio1}) is predictable, i.e., the AWGN-free channel response to any arbitrary input $x_{\mbox{\scriptsize{dd}}}[k,l]$ can be accurately predicted to be $h_{\mbox{\scriptsize{eff}}}[k,l] *_{\sigma} \, x_{\mbox{\scriptsize{dd}}}[k,l]$.

\begin{figure*}
    \centering
    \subfloat[]
    {\includegraphics[width=0.33\linewidth]{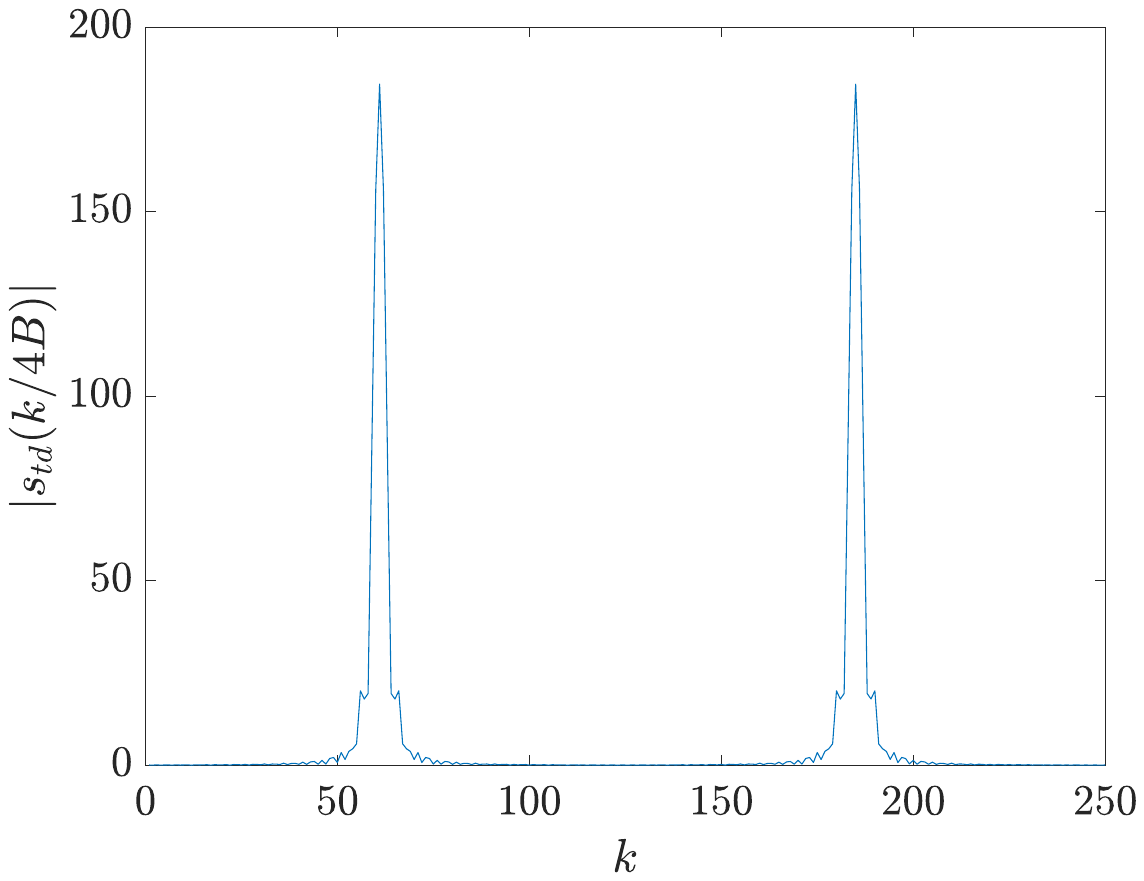}\label{pp}}
    \hfill
    \subfloat[]
    {\includegraphics[width=0.33\linewidth]{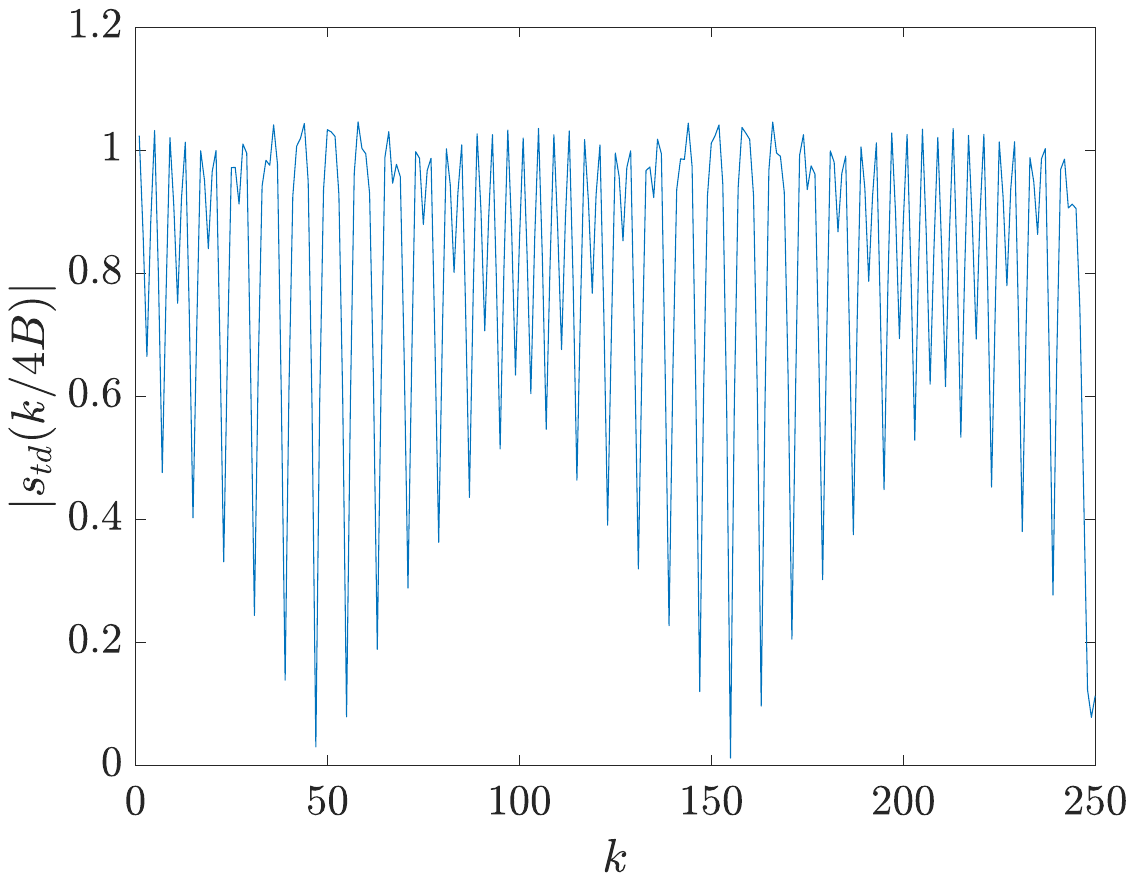}\label{sp}}
    \hfill
    \subfloat[]
    {\includegraphics[width=0.33\linewidth]{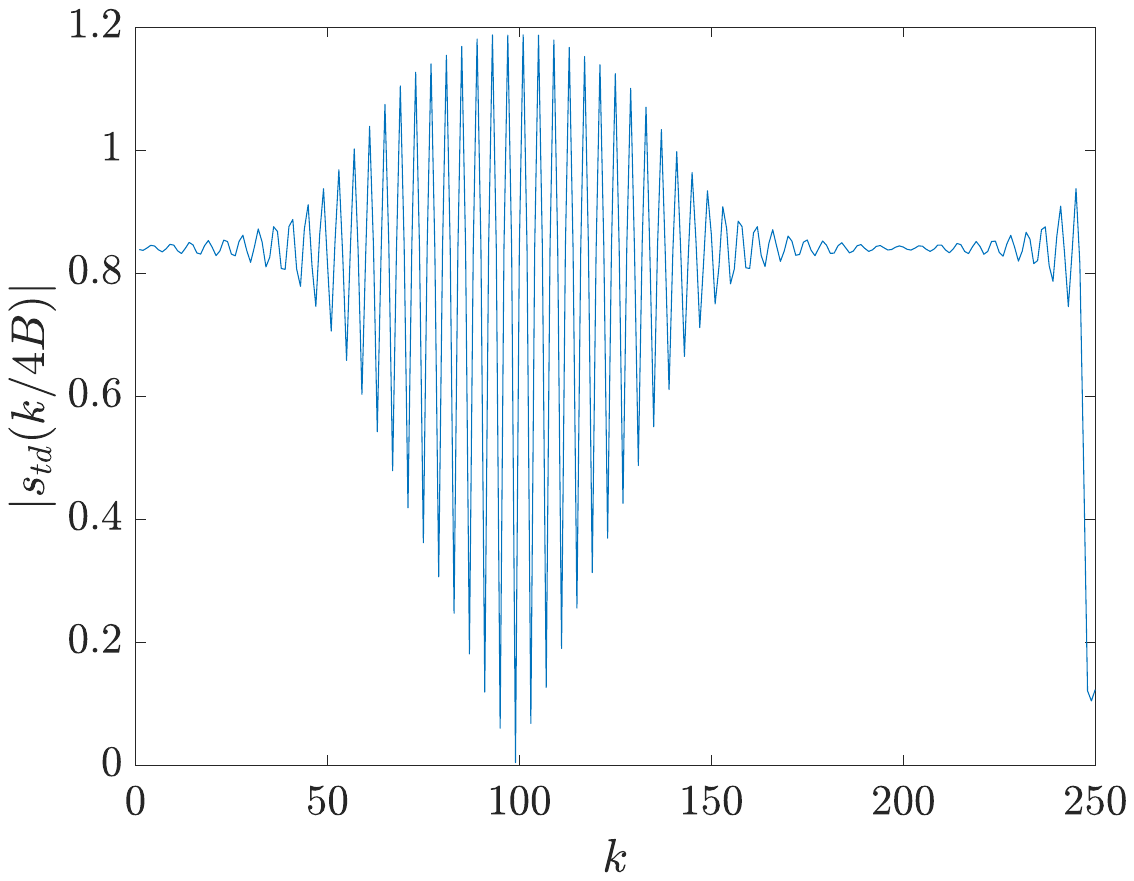}\label{zc}}
    \caption{TD realization of (a) point pulsone located at $(k_p, l_p) = ((M + 1)/2, (N + 1)/2)$, (b) Zak-OTFS spread pilot with slope $q=3$, and (c) ZC pilot with root $u=23$. In each case, $M = 31, N = 37$ for a Zak-OTFS grid with Doppler period $\nu_p 30$ KHz. RRC pulse shaping with $\beta_\tau = \beta_\nu = 0.6$. Only part of the TD pulsone is shown, with samples taken every $1/4B$ seconds where the bandwidth $B = M\nu_p = 930$ KHz.}
    \label{fig:td_realization}
\end{figure*}

When the crystallization conditions hold, there is no aliasing in the DD domain, and the effective DD domain channel filter taps can simply be read off from the response to a single Zak-OTFS pilot. What then are the practical limitations on pilot design? Pulse trains exhibit a high peak-to-average-power ratio (PAPR), hence we need high-power linear amplifiers, and unfortunately, these are power-inefficient. We now describe two methods of constructing spread pilots. Fig. \ref{fig:td_realization} compares the TD realization of the point pulsone with those of the spread pilots.

\textbf{Zak-OTFS Spread Pilots:} First we observe that a discrete quasi-periodic DD domain signal is periodic along both delay and Doppler axes with period $MN$. We define a \textit{discrete DD domain filter} to be a discrete periodic DD domain function with period $MN$ along both delay and Doppler axes. We obtain a spread pilot $x_{dd}^{q}[k, l]$ by applying a discrete spreading filter $w_{q}[k, l]$ to the DD domain signal $x_{p, d d}[k, l]$ localized at the point $p=(0,0)$ in the information lattice (see (\ref{eqnexprxpdd1})). Following \cite{ISAC} we consider chirp filters
\begin{equation}
w_{q}[k, l]=\xi_{M N}^{q k^{2}+q l^{2}}\ 
\end{equation}
where $\xi_{M N}$ is a primitive $MN^{\text {th }}$ root of unity and the slope $q$ is coprime to $M$ and $N$. The spread pilot is given by
\begin{align}
x_{d d}^{q}[k, l]  &= w_{q}[k, l] *_{\sigma} x_{p, d d}[k, l] \nonumber \\
&=\sum_{n=0}^{N-1} \sum_{m=0}^{M-1} w_{q}[k-n M, l-m N] \xi_{N}^{n l}
\end{align}
where $\xi_{N}$ is a primitive $N^{\text {th }}$ root of unity (see \cite{ISAC}, Appendix I ). Write $q=a M+b N$ with $0 \hspace{-1mm}\leqslant \hspace{-1mm}a \hspace{-1mm}\leqslant \hspace{-1mm}N\hspace{-1mm}-\hspace{-1mm}1$ and 
$0 \hspace{-1mm}\leqslant \hspace{-1mm}b \hspace{-1mm}\leqslant \hspace{-1mm} M\hspace{-1mm}-\hspace{-1mm}1$. We have derived the following closed form expression for $x_{d d}^{q}[k, l]$ that greatly simplifies calculation of self-ambiguity and cross-ambiguity functions (cf. \cite{ISAC}, Appendix I):
\begin{equation}
x_{d d}^{q}[k, l]= \frac{1}{\sqrt{M N}} \xi_{N}^{a M l^{2}} \xi_{M}^{-b N k^{2}} \xi_{N}^{-\frac{l^{2}}{4 a M}} \xi_{N}^{k l}
\label{dd_representation}
\end{equation}
where $\xi_{M}, \xi_{N}$ are primitive $M^{\text {th }}$ and $N^{\text {th }}$ roots of unity. The derivation is similar to that of the closed form expression for Zadoff-Chu spread pilots and we omit the details.

The self-ambiguity function $A_{q}[k, l]$ is given by 
\begin{align}
A_{q}[k, l]=\sum_{k^{\prime}=0}^{M-1} \sum_{l^{\prime}=0}^{N-1} x_{d d}^{q}\left[k^{\prime}, l^{\prime}\right] \bar{x}_{d d}^{q}\left[k^{\prime}-k, l^{\prime}-l\right] \xi_{M N}^{-l\left(k^{\prime}-k\right)}
\end{align}
\noindent We have used our closed form expression for $x_{\text {dd }}^{q}[k, l]$ to show that $A_{q}[k, l]$ is supported on a lattice $\Lambda_{q}$ (cf. \cite{ISAC}, Appendix I). Given odd integers $M, N$ and $q$ relatively prime to both $M$ and $N$

\begin{equation}
\left|A_q[k, l]\right|=\left\{\begin{array}{ll}
1, & \text { if } k-\left(\frac{1}{2 a M} -2 a M\right) l \equiv 0 \bmod N \\
 & \text { and } l-2 b N k \equiv 0 \bmod M \\
0, & \text{otherwise}
\end{array}\right.
\end{equation}
The derivation is similar to that of the self-ambiguity function of Zadoff-Chu spread pilots (Theorem 2) and we omit the details since the result appears in \cite{ISAC}.

The lattice $\Lambda_{q}$ is obtained from the period lattice $\Lambda_{p}$ by applying a non-singular linear transformation. When the channel satisfies the crystallization condition with respect to the lattice $\Lambda_{q}$, the effective DD domain filter taps can be read off from the response to a single spread pilot. When the channel satisfies the crystallization condition with respect to the period lattice $\Lambda_{p}$, then given the I/O response at one point in a Zak-OTFS subframe it is possible to predict the response at all points in the subframe.

The cross-ambiguity $A_{q, p}[k, l]$ between spread pilots $x_{d d}^{q}$
 and $x_{d d}^{p}$ is given by
\begin{equation}
A_{q, p}[k, l]=\sum_{k^{\prime}=0}^{M-1} \sum_{l^{\prime}=0}^{N-1} x_{d d}^{q}\left[k^{\prime}, l^{\prime}\right] \bar{x}_{d d}^{p}\left[k^{\prime}-k, l^{\prime}-l\right] \xi_{M N}^{l\left(k^{\prime}-k\right)}
\end{equation}
The cross-ambiguity $A_{q, p}[k, l]$ measures the angle between $x_{d d}^{q}$ and $x_{d d}^{p}$ after a delay shift $k$ and a Doppler shift $l$. We have used our closed form \eqref{dd_representation} to show that $A_{q, p}[k, l]$ takes the form
\begin{equation}
C_{k, l} A_{q, p}[k, l]=\frac{1}{M N} \left(\sum_{l^{\prime}=0}^{N-1} \xi_{N}^{f l'^{ 2}}\right)\left(\sum_{k^{\prime}=0}^{M-1} \xi_{M}^{a k'^{ 2}} \xi_{M N}^{b k^{\prime}}\right)
\end{equation}
where $C_{k, l}$ is a complex phase depending only on $k, l$, the residue $f \not\equiv  0(\bmod N)$ and the residues $a, b \not\equiv  0(\bmod M)$. We have shown that
\begin{equation}
\left|\left(\sum_{l^{\prime}=0}^{N-1} \xi_{N}^{f l'^{ 2}}\right)\right|=\sqrt{N} \text { and }\left|\left(\sum_{k^{\prime}=0}^{M-1} \xi_{M}^{a k'^{ 2}} \xi_{M N}^{b k^{\prime}}\right)\right|=\sqrt{M}
\end{equation}
so that $x_{d d}^{q}$ and $x_{d d}^{p}$ are unbiased after arbitrary delay and Doppler shifts. Again, the derivation is similar to that of the cross-ambiguity of Zadoff-Chu spread pilots (Theorem 3), and we omit the details since the result appears in \cite{ISAC}.

\textbf{Zadoff-Chu Spread Pilots:} A Zadoff-Chu (ZC) sequence $X_{u}[n]$ with \textit{root} $u$ and composite length $MN$ is given by
\begin{equation}
X_{u}[n]=\xi_{M N}^{-u n(n+1)/2} \quad \text { where } n=0,1, \ldots, M N-1
\end{equation}
We take $M, N$ to be odd integers, with $M$ coprime to $N$. The ZC sequence $X_{u}[n]$ determines an $MN$-periodic sequence $\tilde{X}_{u}[n]$ that satisfies $\widetilde{X}_{u}[n+s M N]=X_{u}[n]$. We apply the Zak transform to give the discrete DD domain pilot
\begin{align}
\sqrt{MN}X_{d d}^{u}[k, l]=\frac{1}{\sqrt{N}} \sum_{p=0}^{N-1} \tilde{X}_{u}\left[k+pM\right] \xi_{N}^{-p l}
\end{align}
Normalization by $\sqrt{MN}$ results in a pilot with unit energy.

We derive our closed form expression for $X_{d d}^{u}[k, l]$ using the following lemma.\\
\textbf{Lemma 1:} Let $N$ be an odd integer, and let $\xi_N$ denote a primitive $N^{\text {th }}$ root of unity. If $a$ is an integer coprime to $N$, then
\begin{align}
    S_a = \sum_{p=0}^{N-1}\xi_N^{ap^2} = C_a\sqrt{N},
\end{align}
where $C_a$ is a complex phase.\\
\textbf{Proof:} We have
\begin{align}
    S_a\bar{S}_a = \sum_{p_1=0}^{N-1}\sum_{p_2=0}^{N-1} \xi_{N}^{a(p_1^2 - p_2^2)}.
\end{align}
Since $N$ is odd, we may change variables by setting $p_3 = p_1+p_2$, $p_4=p_1-p_2$ to obtain
\begin{align}
    S_a\bar{S}_a = \sum_{p_3=0}^{N-1}\sum_{p_4=0}^{N-1} \xi_{N}^{ap_3p_4} = N.
\end{align}
The following Theorem gives the closed form expression for $X_{d d}^{u}[k, l]$.\hfill$\blacksquare$\\\\
\textbf{Theorem 1:} Let $M$, $N$ be odd integers with $M$ coprime to $N$. Then to within a complex phase independent of $k$ and $l$
\begin{align}
    X_{d d}^{u}[k, l] = \frac{1}{\sqrt{MN}}\xi_{MN}^{-\frac{uk(k+1)}{2}}\xi_N^{\frac{(u(2k+1)+2l)^2}{8uM}}.
    \label{theroem1_eq1}
\end{align}
\textbf{Proof:} We have
\begin{align}
    \sqrt{MN}X_{d d}^{u}[k, l] = \frac{1}{\sqrt{N}}\sum_{p=0}^{N-1}\bar{X}_u[k+pM]\xi_N^{-pl},
    \label{theroem1_eq2}
\end{align}
where
\begin{align}
    \bar{X}_u[k+pM] &= \xi_{MN}^{-\frac{u(k+pM)(k+pM+1)}{2}} \nonumber \\
    &= \xi_{MN}^{-\frac{uk(k+1)}{2}}\xi_{N}^{-\frac{u(2kp+p^2M+p)}{2}}
    \label{theroem1_eq3}
\end{align}
Now
\begin{align}
    \frac{1}{\sqrt{N}}&\sum_{p=0}^{N-1}\xi_N^{-\frac{u(2kp+p^2M+p)}{2}}\xi_N^{-pl} \nonumber \\
    &\hspace{-3mm}=\frac{1}{\sqrt{N}}\sum_{p=0}^{N-1}\xi_{N}^{-\frac{uM}{2}\left(p^2+\frac{2kp}{M}+\frac{p}{M}+\frac{2lp}{uM}\right)} \nonumber \\
    &\hspace{-3mm}=\frac{1}{\sqrt{N}}\xi_N^{\frac{uM}{2}\left(\frac{k}{M}+\frac{1}{2M}+\frac{l}{uM}\right)^2}\sum_{p=0}^{N-1}\xi_N^{-\frac{uM}{2}\left(p+\left(\frac{k}{M}+\frac{1}{2M}+\frac{l}{uM}\right)\right)^2}.
    \label{theroem1_eq4}
\end{align}
It follows from Lemma 1 that
\begin{align}
    \frac{1}{\sqrt{N}}\sum_{p=0}^{N-1}\xi_N^{-\frac{uM}{2}\left(p+\left(\frac{k}{M}+\frac{1}{2M}+\frac{l}{uM}\right)\right)^2} = C,
    \label{theroem1_eq5}
\end{align}
where $C$ is a complex phase independent of $k$ and $l$. Finally
\begin{align}
    \frac{uM}{2}\left(\frac{k}{M}+\frac{1}{2M}+\frac{l}{uM}\right)^2 = \frac{(u(2k+1)+2l)^2}{8uM}
    \label{theroem1_eq6}
\end{align}
and we combine \eqref{theroem1_eq2} - \eqref{theroem1_eq6} to obtain \eqref{theroem1_eq1}. \hfill$\blacksquare$\\
The self-ambiguity function $A_u[k, l]$ is given by
\begin{align}
    A_u[k, l] = \sum_{k'=0}^{M-1}\sum_{l'=0}^{N-1}X_{d d}^{u}[k', l']\bar{X}_{d d}^{u}[k'-k, l'-l]\xi_{MN}^{-l(k'-k)}.
    \label{self_ambiguity}
\end{align}
We now show that the self-ambiguity function of the ZC pilot with root $u$ is supported on the line $l= - uk\mod MN$.\\\\
\textbf{Theorem 2:} Let $M, N$ be odd integers with $M$ coprime to $N$. Then
\begin{align}
    \vert A_u[k, l]\vert = \begin{cases}
        1, \quad \text{if } l = - uk\mod MN \\
        0, \quad \text{otherwise}
    \end{cases}.
    \label{theorem2_eq1}
\end{align}
\textbf{Proof:} We have
\begin{align}
    X_{d d}^{u}[k', l'] = \frac{1}{\sqrt{MN}}\xi_{MN}^{-\frac{uk'(k'+1)}{2}}\xi_{N}^{\frac{\left(u(2k'+1)+2l'\right)^2}{8uM}}
    \label{theorem2_eq2}
\end{align}
and
\begin{align}
    \bar{X}_{d d}^{u}[k'-k, l'-l] = \frac{1}{\sqrt{MN}}&\xi_{MN}^{\frac{u(k'-k)(k'-k+1)}{2}}\times\nonumber\\
    &\xi_{N}^{-\frac{\left(u(2(k'-k)+1)+2(l'-l)\right)^2}{8uM}}.
    \label{theorem2_eq3}
\end{align}
The exponent
\begin{align}
    \frac{u(k'-k)(k'-k+1)}{2} -\frac{uk'(k'+1)}{2} = \frac{uk^2 - 2ukk' - uk}{2}.
    \label{theorem2_eq4}
\end{align}
The exponent
\begin{align}
    &\left(u(2k'+1)+2l'\right)^2 - \left(u(2(k'-k)+1)+2(l'-l)\right)^2 \nonumber \\
    &=(u(2k'+1)+2l'+u(2(k'-k)+1) + 2(l'-l)) \times \nonumber \\
    &\hspace{11mm}(u(2k'+1)+2l' - (u(2(k'-k)+1)+2(l'-l))) \nonumber \\
    &=2(l+uk)(4uk'-2uk+2u+4l'-2l).
    \label{theorem2_eq5}
\end{align}
We substitute \eqref{theorem2_eq4}, \eqref{theorem2_eq5} in \eqref{self_ambiguity}, observing that all terms involving $k$ and $l$, but neither $k'$ nor $l'$, just contribute an overall phase $C$. We obtain
\begin{align}
    A_u[k, l] = \frac{C}{MN}\sum_{k'=0}^{M-1}\sum_{l'=0}^{N-1} \xi_{MN}^{-ukk'}\xi_N^{\frac{(l+uk)(uk'+l')}{uM}}\xi_{MN}^{-lk'}.
\end{align}
The inner sum
\begin{align}
    \sum_{l'=0}^{N-1} \xi_N^{\frac{(l+uk)l'}{uM}} = \begin{cases}
        N, \quad \text{if } l = -uk\mod N \\
        0, \quad \text{ otherwise}
    \end{cases}.
\end{align}
Hence
\begin{align}
    A_u[k, l] \hspace{-1mm} = \hspace{-1mm} \frac{C}{M}\sum_{k'=0}^{M-1}\hspace{-1mm}\xi_{MN}^{-(uk+l)k'}\hspace{-2mm} = \hspace{-1mm}\begin{cases}
        C, \  \text{if } l = -uk \mod M  \\
        0, \  \text{ otherwise}
    \end{cases}\hspace{-2mm}.
\end{align}
Hence $A_{u}[k, l]$ is supported on the line $l = -uk \mod MN$. \\
{\color{white} a}$\hfill\blacksquare$\\
We use the following lemma to estimate the size of the cross-ambiguity $A_{u, w}[k, l]$ between ZC pilots $X_{d d}^{u}$ and $X_{d d}^{w}$. \\\\
\textbf{Theorem 3:} Let $M, N$ be odd integers. If $X_{d d}^{u}, X_{d d}^{w}$ are distinct ZC pilots, then the cross-ambiguity $A_{u, w}[k, l]$ satisfies
\begin{align}
    A_{u, w}[k, l] = \frac{C_{k, l}}{\sqrt{MN}},
\end{align}
where $C_{k, l}$ is a complex phase.\\\\
\textbf{Proof:} The cross-ambiguity $A_{u, w}[k, l]$ is given by
\begin{align}
    A_{u, w}[k, l] = \sum_{k'=0}^{M-1} \sum_{l'=0}^{N-1} X_{d d}^{u}[k', l'] \bar{X}_{d d}^{w}[k'-k, l'-l]\xi_{MN}^{-l(k'-k)}.
    \label{theorem3_eq1}
\end{align}
It follows from Theorem 1 that
\begin{align}
    X_{d d}^{u}[k', l'] = \frac{1}{\sqrt{MN}}\xi_{MN}^{-\frac{uk'(k'+1)}{2}}\xi_N^{\frac{(u(2k'+1)+2l')^2}{8uM}}
    \label{theorem3_eq2}
\end{align}
and
\begin{align}
    \bar{X}_{d d}^{w}[k'-k, l'-l] = \frac{1}{\sqrt{MN}}&\xi_{MN}^{\frac{w(k'-k)(k'-k+1)}{2}} \times \nonumber \\
    &\xi_N^{-\frac{(w(2(k'-k)+1)+2(l'-l))^2}{8wM}}.
    \label{theorem3_eq3}
\end{align}
We first show that there are no cross terms $k'l'$ in the cross-ambiguity $A_{u, w}[k, l]$. This implies that $A_{u, w}[k, l]$ factors as the product of a sum on $k'$ and a sum on $l'$. Again, we note that all terms involving $k$ and $l$ but neither $k'$ nor $l'$ just contribute an overall phase to $A_{u, w}[k, l]$. The cross terms $k'l'$ appear in the product
\begin{align}
    \xi_N^{\frac{(u(2k'+1)+2l')^2}{8uM}}\xi_N^{-\frac{(w(2(k'-k)+1)+2(l'-l))^2}{8wM}} = \xi_N^{\frac{\mathcal{P}}{8uwM}}.
    \label{theorem3_eq4}
\end{align}
We have
\begin{align}
    \mathcal{P} &= w(u(2k'+1)+2l')^2 - u(w(2(k'-k)+1)+2(l'-l))^2 \nonumber \\
    &= w(u(2k'+1)+2l')^2 - u(w(2k'+1) + 2(l'-l -wk))^2 \nonumber \\
    &= w(u^2(2k'+1)^2 + 4l'u(2k'+1) + 4l'^2) - \nonumber \\
    & \hspace{-3mm}u(w^2(2k'+1)^2 + 4w(2k'+1)(l'-l-wk) + 4(l'-l-wk)^2)  \nonumber \\
    &= w(u^2(2k'+1)^2+4l'^2)-\nonumber \\
    & u(w^2(2k'+1)^2 - 4w(2k'+1)(l+wk) + 4(l'-l-wk)^2).
    \label{theorem3_eq5}
\end{align}
There are no cross terms $k'l'$. So the cross-ambiguity $A_{u, w}[k, l]$ factors as a product of two sums. The sum on $l'$ is
\begin{align}
    S_1 = \sum_{l'=0}^{N-1} \xi_{N}^{\frac{4(w-u)l'^2 + 8ul'(l+wk)}{8uwM}}.
    \label{theorem3_eq_6}
\end{align}
We complete the square in the exponent modulo $N$, then apply Lemma 1 to obtain $S_1 = C_1\sqrt{N}$, were $C_1$ is a complex phase.

Each term in the sum on $k'$ is the product $\xi_{MN}^{\mathcal{G}_1}\xi_{N}^{\mathcal{G}_2}$ of two terms.
We have 
\begin{align}
    \mathcal{G}_1 &= \frac{-uk'(k'+1) + wk'(k'-k+1) - wkk'}{2} \nonumber \\
    &= (w-u)\frac{k'^2 + k'}{2} - wkk'
    \label{theorem3_eq_7}
\end{align}
and
\begin{align}
    \mathcal{G}_2 &= \frac{w(u^2(4k'^2+4k')) - u(w^2(4k'^2+4k') - 8wk'(l+wk))}{8uwM} \nonumber \\
    &= \frac{u(k'^2+k') - w(k'^2+k') + 2k'(l+wk)}{2M} \nonumber \\
    &= \frac{(u-w)\frac{(k'^2 + k')}{2} + lk' +wkk'}{M}.
\end{align}
Observe that $\xi_N^{\frac{1}{M}}$ and $\xi_{MN}$ are the $M^{\text {th }}$ roots of $\xi_N$ so they differ by an $M^{\text {th }}$ root of unity which we denote $\xi_M$. The sum on $k'$ becomes
\begin{align}
    \sum_{k'=0}^{M-1} \xi_{MN}^{\mathcal{G}_1 + \mathcal{G}_2 -lk'} \xi_{M}^{\mathcal{G}_2}
\end{align}
and we observe that $\mathcal{G}_1 + \mathcal{G}_2 -lk' = 0$. The sum on $k'$ reduces to
\begin{align}
    S_2 = \sum_{k'=0}^{M-1} \xi_M^{(u-w)\frac{(k'^2 + k')}{2} + lk' +wkk'}.
\end{align}
We complete the square in the exponent modulo $M$, then apply Lemma 1 to obtain $S_2 = C_2\sqrt{M}$, where $C_2$ is a complex phase. \hfill$\blacksquare$\\

\begin{figure}
    \centering
    \includegraphics[width=\linewidth]{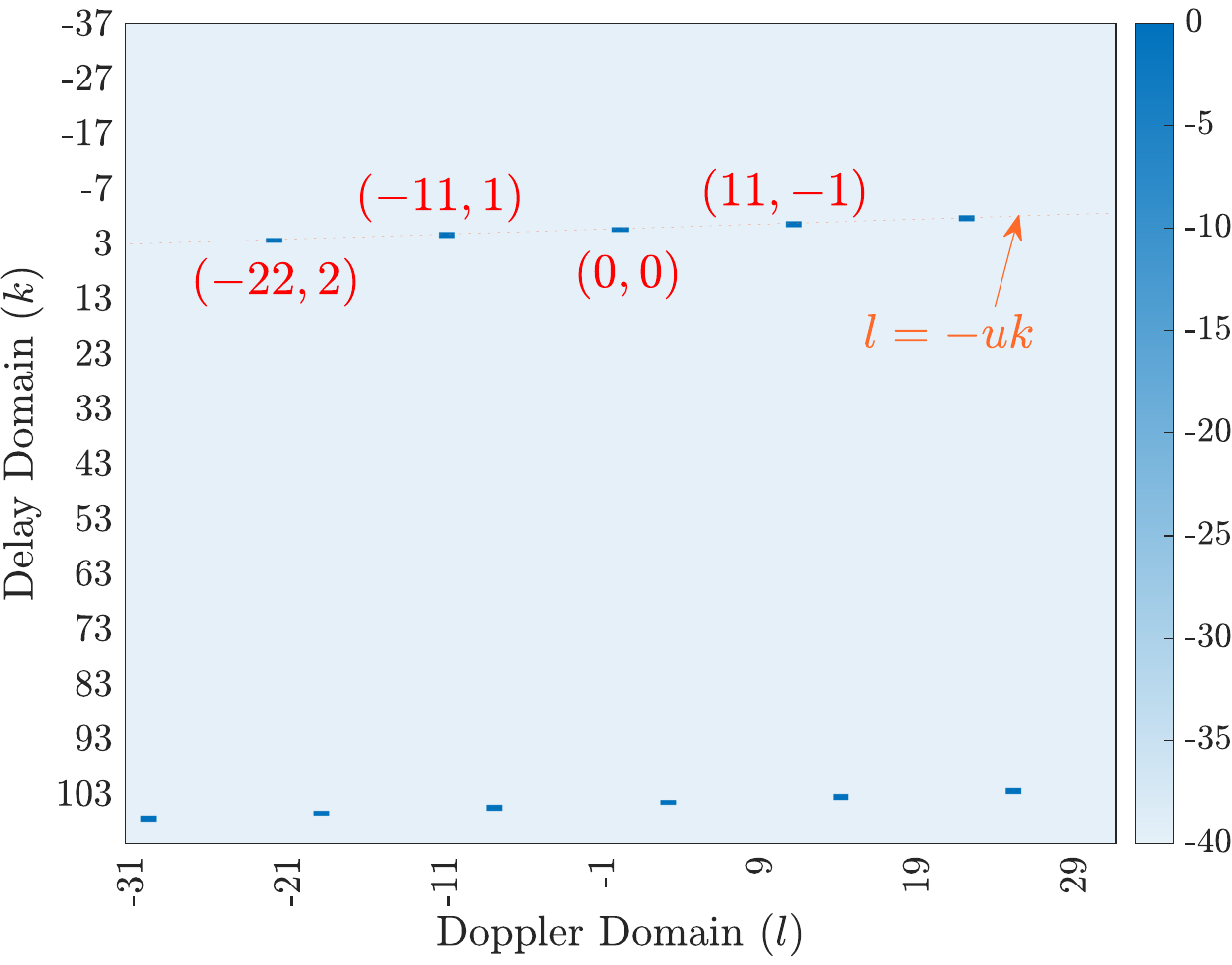}
    \caption{Blue dots mark the support of the self-ambiguity function of the Zadoff-Chu spread pilot with $u = 11$, $M = 31, N=37$. The self-ambiguity function is supported on the line $l=-11k$.}
    \label{fig:self_ambiguity}
\end{figure}

\begin{figure}
    \centering
    \subfloat[]
    {\includegraphics[width=0.5\linewidth]{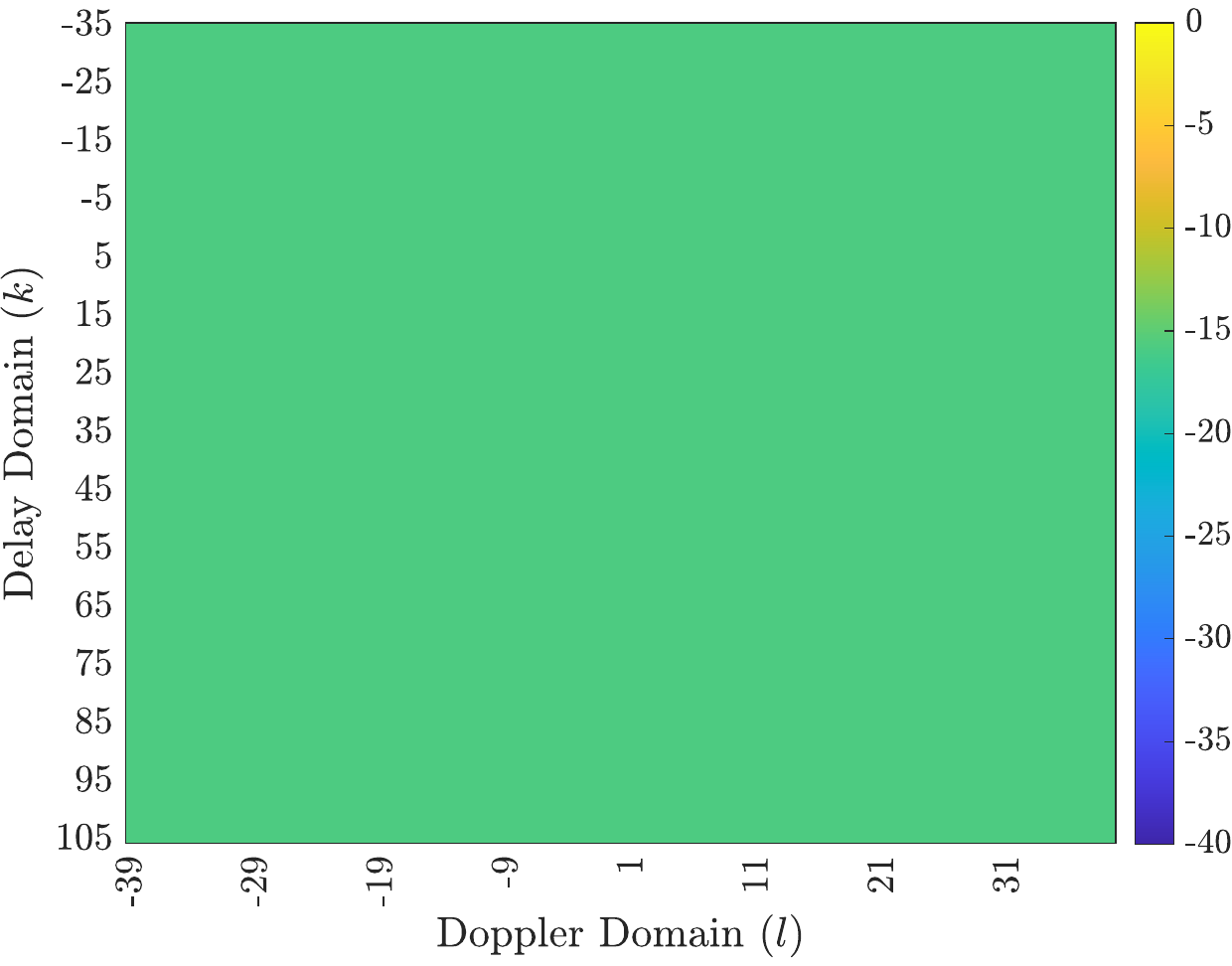}\label{m_odd_n_odd}}
    \hfill
    \subfloat[]
    {\includegraphics[width=0.5\linewidth]{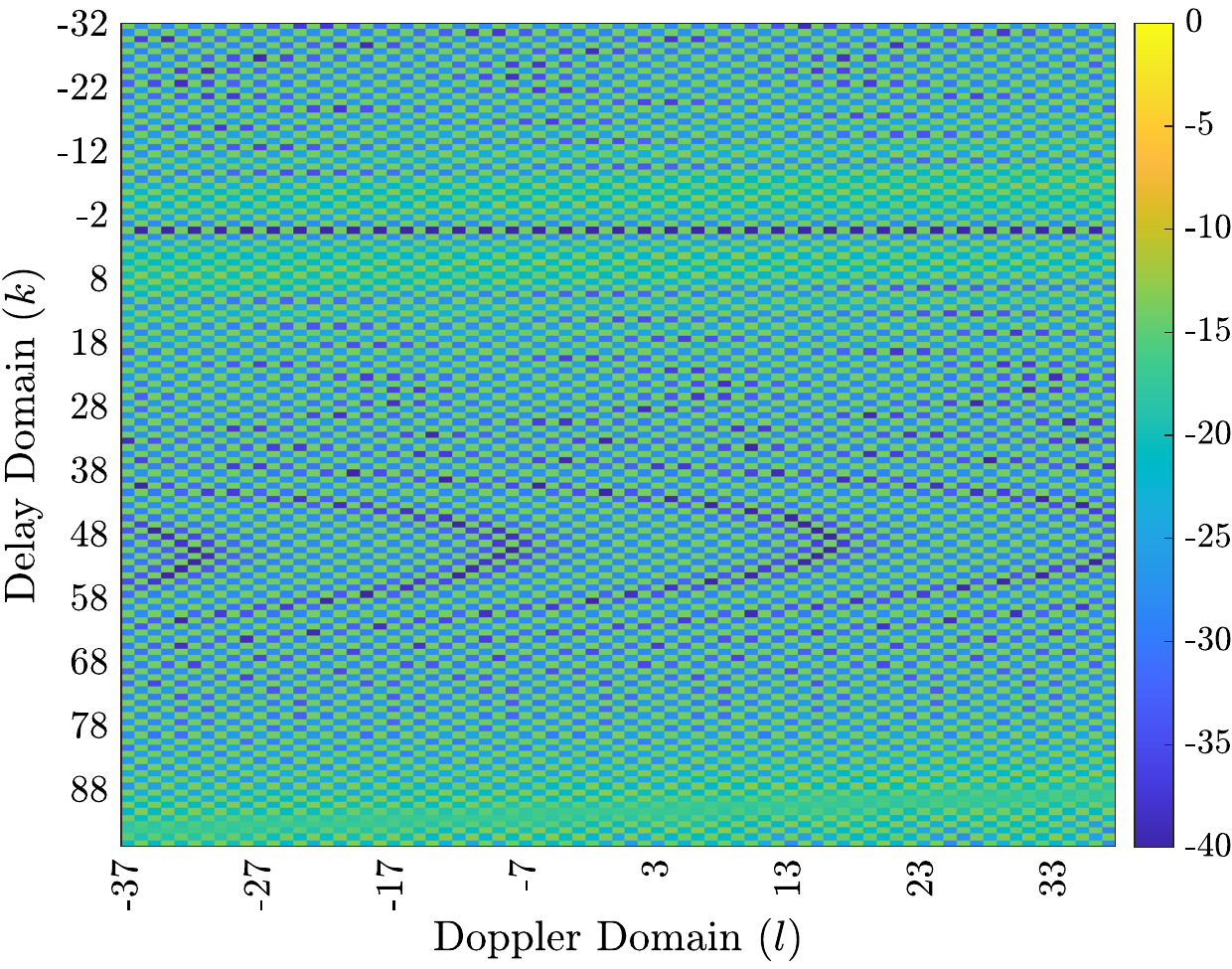}\label{m_even_n_odd}}
    \caption{Magnitude of the cross-ambiguity function. (a) Here $M = 35, N = 39$ are odd and $u = 11, w = 13$ and the cross-ambiguity is flat ($\vert A_{u, w}[k, l] \vert = 1/\sqrt{MN}$). (b) Here $M=32$ is even, $N=37, u = 11, w = 13$ and the cross ambiguity is no longer flat.}
    \label{fig:cross_ambiguity}
\end{figure}

\section{Data Transmission}
\label{sec:datatransmission}
\begin{table}[t]
    \centering
    \caption{Power-delay profile of Veh-A channel model}
    \begin{tabular}{|c|c|c|c|c|c|c|}
         \hline
         Path number ($i$) & 1 & 2 & 3 & 4 & 5 & 6 \\
         \hline
         $\tau_i (\mu s)$ & 0 & 0.31 & 0.71 & 1.09 & 1.73 & 2.51 \\
         \hline
         Relative power ($p_i$) dB & 0 & -1 & -9 & -10 & -15 & -20 \\
         \hline
    \end{tabular}
    \label{tab:veh_a}
\end{table}

In this Section, we measure uncoded 4-QAM BER performance of Zak-OTFS for root raised cosine pulse. Throughout this Section, we fix $M = 31, N = 37$, $\nu_p = 30$ KHz (unless stated otherwise), and we fix the pilot location $(kp, lp) = (M/2, N/2)$ for the Zak-OTFS spread pilot. The bandwidth $B = M\nu_p = 930$ KHz and the duration $T = N\tau_p = 1.23$ ms. Also throughout this Section,
we consider the six-path Veh-A channel model \cite{veh_a} with power-delay profile listed in Table \ref{tab:veh_a}. The relative power $p_i$ of the $i$-th path is the ratio (in dB) of the mean squared value of the $i$-th tap to that of the first tap ($pi = \mathbb{E}[\vert h_i\vert^2]/ \mathbb{E}[\vert h_1\vert^2], i = 1, 2, \cdots , 6$). We then normalize these mean squared values
so that $\sum_{i=1}^6 \mathbb{E}[\vert h_i\vert^2] = 1$. The Doppler shift $\nu_i$ induced by
the $i$-th channel path is modeled as $\nu_{\max} \cos(\theta_i)$ where $\nu_{\max}$ is the maximum Doppler shift and $\theta_i, i = 1, 2, \cdots , 6$ are modeled as i.i.d. random variables distributed uniformly in $[0 , 2\pi)$.

We follow \cite{ISAC} in estimating the taps of the discrete effective channel filter $h_{\mbox{\scriptsize{eff}}}[k,l]$ from the discrete cross-ambiguity of the received DD signal and the transmitted ZC spread pilot signal. The self-ambiguity function $A_u[k, l]$ of the ZC spread pilot is supported on pairs $(k, l)$ that satisfy $l=-uk \mod MN$. The support $S(k, l)$ of the discrete effective channel filter $h_{\mbox{\scriptsize{eff}}}[k,l]$ at a pair $(k_0, l_0)$ satisfying $l_0=-uk_0 \mod MN$ is given by 
\begin{align}
    S(k_0, l_0) = \{(k, l) \mid h_{\mbox{\scriptsize{eff}}}[k-k_0,l-l_0] \neq 0\}.
\end{align}
The crystallization condition eliminates aliasing in the DD domain by requiring that the sets $S(k_0, l_0)$ are disjoint as $(k_0, l_0)$ ranges over pairs satisfying $l_0=-uk_0 \mod MN$. When the crystallization condition holds we are able to read off the taps of the discrete effective channel filter $h_{\mbox{\scriptsize{eff}}}[k,l]$ from the response to a ZC spread pilot signal (see Section \ref{sec:pilot_design}, and see \cite{OTFS2Paper2}, Section II-D, and \cite{ISAC}, Section V-B for more details).

\begin{figure}
    \centering
    \includegraphics[width=\linewidth]{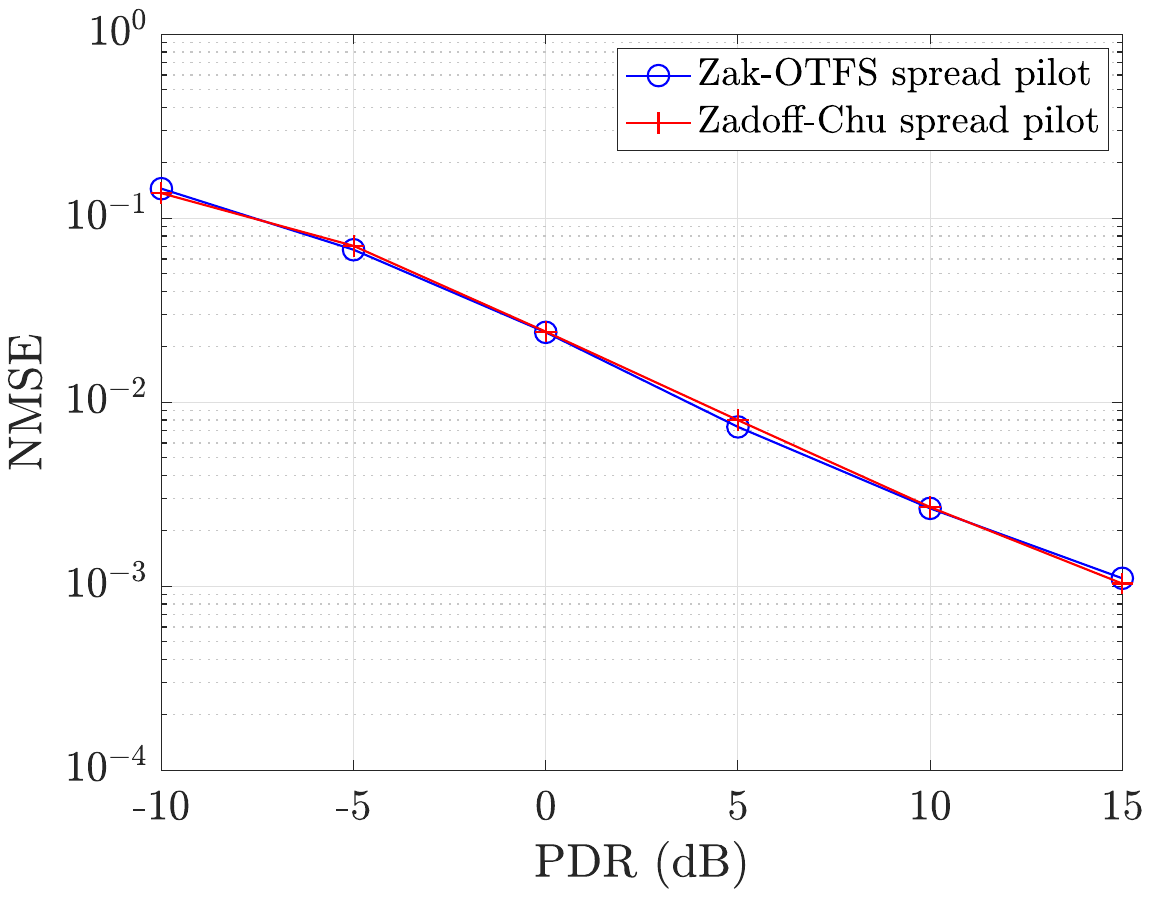}
    \caption{ Normalized mean squared error (NMSE) as a function of increasing PDR (cf. \cite{ISAC}, Fig. 23) for a Zak-OTFS spread pilot and a ZC spread pilot. $M=31, N=37$ with Doppler period $\nu_p = 30$ KHz. Zak-OTFS spreadpilot defined by a $MN$ periodic chirp filter with slope parameter $q=3$. ZC spread pilot with root $u=11$. Veh-A channel, RRC pulse shaping filter ($\beta_\tau = \beta_\nu = 0.6$), data SNR $\rho_d = 25$ dB, $\nu_{\max} = 815$ Hz.}
    \label{fig:nmse_vs_pdr}
\end{figure}

We consider transmission of pilot and data within a single Zak-OTFS subframe, with \textit{no division of radio resources} between sensing the effective channel on the period lattice and data transmission. Data is carried by Zak-OTFS carriers (point pulsones), the data signal interferes with the ZC spread pilot, and this interference adversely affects the estimation of certain discrete effective channel filter taps. After estimating the effective channel, we are able to estimate the received spread pilot, subtract this estimate from the received signal, and recover the data. The residual pilot (after cancellation) interferes with data transmission. Figure \ref{fig:nmse_vs_pdr} illustrates how the significance of this interference depends on the ratio of pilot to data power (PDR). There is essentially no difference between sensing the effective channel with spread Zak-OTFS pilots \cite{ISAC} and sensing the effective channel with ZC spread pilots. 

\begin{figure}
    \centering
    \includegraphics[width=\linewidth]{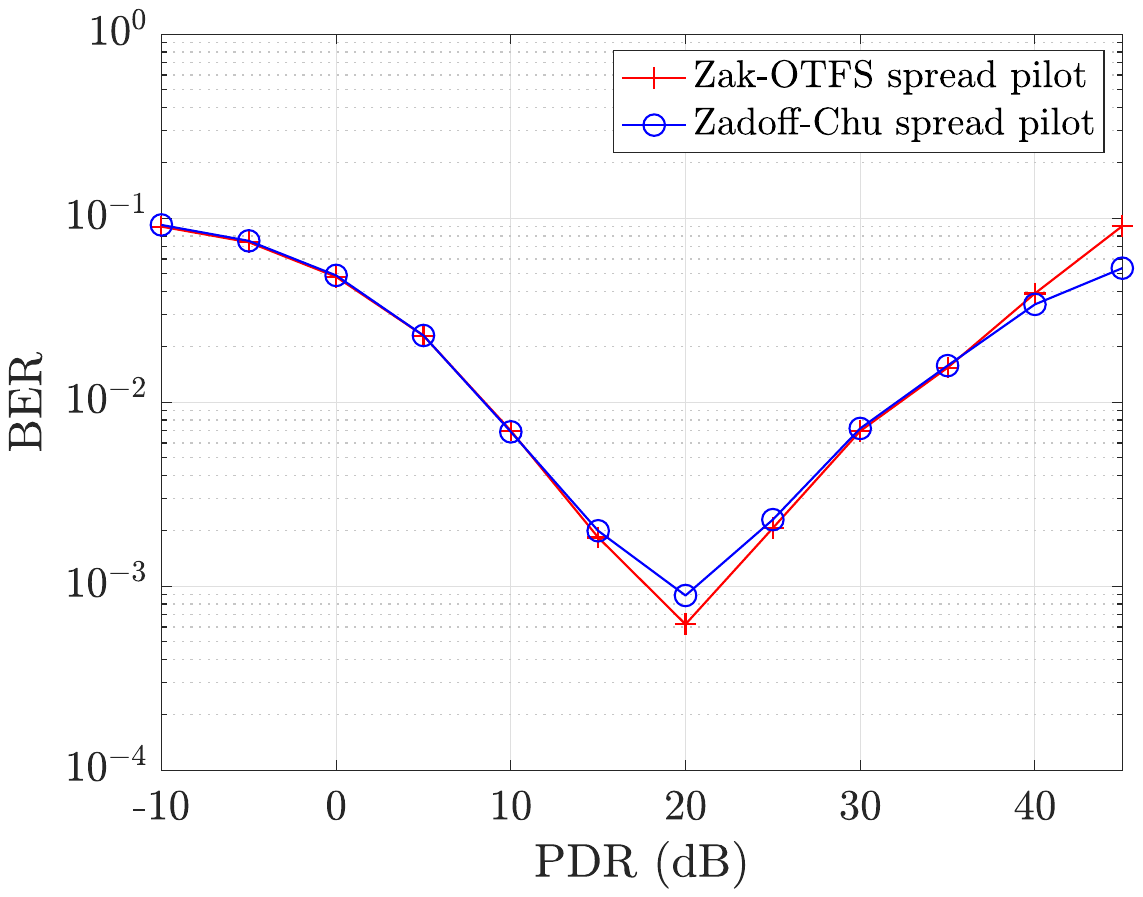}
    \caption{BER performance for uncoded 4-QAM as a function of increasing PDR (cf. \cite{ISAC}, Fig. 24) for a Zak-OTFS spread pilot and a ZC spread pilot. $M=31, N=37$ with Doppler period $\nu_p = 30$ KHz. Zak-OTFS spreadpilot defined by a MN periodic chirp filter with slope parameter $q=3$. ZC spread pilot with root $u=11$. Veh-A channel, RRC pulse shaping filter ($\beta_\tau = \beta_\nu = 0.6$), data SNR $\rho_d = 25$ dB, $\nu_{\max} = 815$ Hz.}
    \label{fig:ber_vs_pdr}
\end{figure}

When sensing the channel with spread Zak-OTFS pilots, the BER performance of uncoded 4-QAM exhibits a ``U-shaped'' dependence on PDR (\cite{ISAC}, Figure 24). When the PDR is less than 20 dB, the channel estimate improves with increasing PDR, leading to improved BER. Beyond 20 dB, interference from the residual pilot (after cancellation) becomes more significant than noise, and PDR degrades with increasing PDR. Figure \ref{fig:ber_vs_pdr} illustrates BER performance of uncoded 4-QAM as a function of increasing PDR. There is essentially no difference in BER performance between sensing the effective channel with spread Zak-OTFS pilots \cite{ISAC} and sensing the effective channel with ZC spread pilots. 

\begin{figure}
    \centering
    \includegraphics[width=0.9\linewidth]{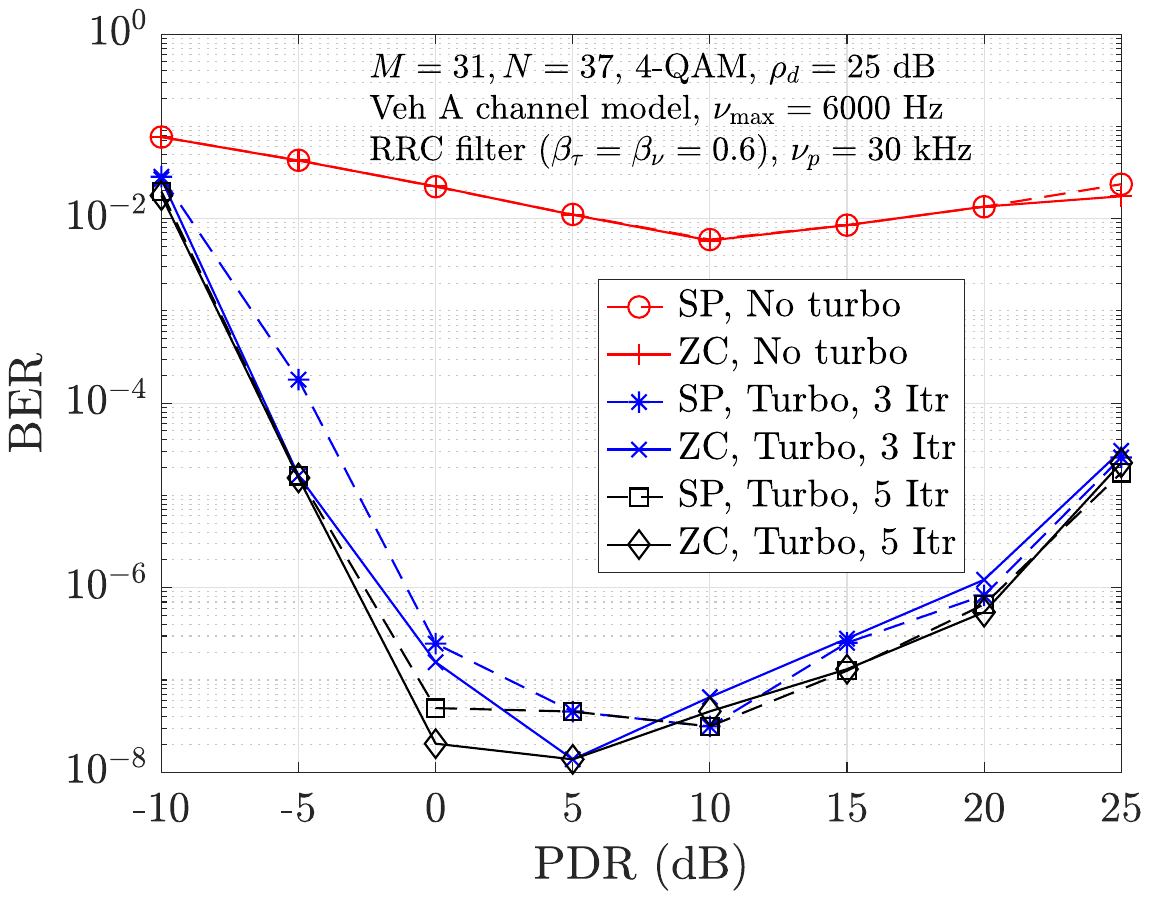}
    \caption{BER performance for uncoded 4-QAM as a function of increasing PDR. Data transmission integrated with sensing the I/O relation in a single Zak-OTFS subframe. ZC spread pilot with root $u=11$. $M=31, N=37$ for Zak-OTFS grid with Doppler period $\nu_p = 30$ KHz. Veh-A channel, RRC pulse shaping filter ($\beta_\tau = \beta_\nu = 0.6$), data SNR $\rho_d = 25$ dB, $\nu_{\max} = 6000$ Hz. Very significant improvement in BER performance from three to five turbo iterations.}
    \label{fig:turbo}
\end{figure}

We have focused on transmission of pilot and data within a single Zak-OTFS subframe since we are interested in minimizing the latency of short packets on the uplink. It is of course possible to improve BER performance by dedicating separate Zak-OTFS subframes to sensing the effective channel and to data transmission. We follow \cite{turbo_paper} in showing that turbo signal processing can close the performance gap between sensing/data transmission in separate Zak-OTFS frames and joint sensing/data transmission in the same Zak-OTFS frame. In the turbo iteration, we take the estimated data, then estimate the received data signal, then improve our estimate for the effective channel by subtracting our estimate for the received data signal from the received signal. Fig. \ref{fig:turbo} shows that with turbo signal processing, there is essentially no difference in BER performance between sensing the effective channel with spread Zak-OTFS pilots \cite{ISAC} and sensing the effective channel with ZC spread pilots. 

In joint sensing and communication, we estimate the effective channel, then estimate the received ZC spread pilot, then recover the data after subtracting our estimate for the received pilot from the received signal. The residual pilot (after cancellation) interferes with data transmission, and we conclude this Section by introducing a turbo iteration that systematically reduces this interference. We take the estimated data, then estimate the received data signal, then improve our estimate for the effective channel by subtracting our estimate for the received data signal from the received signal. Figure \ref{fig:turbo} shows that five turbo iterations suffice to approach the BER performance obtained by dedicating a Zak-OTFS subframe to sensing the I/O relation that is separate from the Zak-OTFS subframe used to transmit data(see \cite{turbo_paper} for a parallel development using Zak-OTFS spread pilots).

\section{Mutliple Preamble Detection}
\label{sec:preamble_detection}
\newcommand{\bigslant}[2]{{\raisebox{.2em}{$#1$}\left/\raisebox{-.2em}{$#2$}\right.}}

\begin{algorithm}
    \caption{The OST algorithm for detecting multiple preambles}
    \label{alg:ost}
    \begin{algorithmic}
        \STATE  \textbf{Input:} An $n \times p$ observation matrix $\mathbf{A}$, the received vector $\mathbf{y} \in \mathbb{C}^{n}$, number of simultaneous users $K$, set of delay and Doppler values $\mathcal{S}$, threshold $\lambda > 0$
        \STATE \textbf{Output:} An estimate of the transmitted preambles
        \STATE  $\mathbf{f} \leftarrow \mathbf{A}^H\mathbf{y}$
        \STATE  $\mathcal{J} \leftarrow \left\{\bigslant{i}{\vert \mathcal{S}\vert} \ \forall \ i \in \{1,2,\ldots,p\} \right\}$
        \STATE $\mathcal{J}_j \leftarrow \left\{i \in \{1,2,\ldots,p\} \mid \bigslant{i}{\vert \mathcal{S}\vert} = j, \ j \in \mathcal{J}\right\}$
        \STATE $\mathcal{J}'\leftarrow \left\{j \mid \sum_{i \in \mathcal{J}_j}\vert \mathbf{f}_i \vert^2 > \lambda\ \ \forall j \in \mathcal{J}\right\}$ such that $\vert\mathcal{J}'\vert = K$
        \STATE $\mathcal{J} \leftarrow \mathcal{J}'$
        \RETURN $\mathcal{J}$
    \end{algorithmic}
\end{algorithm}

In this section, we present a Zak-OTFS-based multiple access scheme for the uplink with $K$ active users when the communication channel is doubly dispersive.
We embrace the 2-step RACH based multiple access considered in \cite{peralta2021two,agostini2024evolution}.
In this scheme, each active user picks a random preamble from a set of preambles, and transmits them simultaneously. 
Each preamble points to a time-frequency resource block for transmission of the user's data.
At the base station, the active preambles are detected and then,
demodulation/decoding is performed in the resource blocks associated with the preamble. 
We propose the use of Zadoff-Chu spread pilots as preambles and the Zak-OTFS based data transmission schemes explained in Section~\ref{sec:datatransmission} for transmitting data in the resource block associated with the preamble.

We focus on the detection of multiple preambles 
since preamble detection is a critical first step in recovering a user's data.
Without detecting the preamble of a user, the pointer to the resources used by the user is lost. 
Not only does this render detection of the user's data impossible, this also causes undetected interference to other users.
Once the preambles are detected data transmission and detection can be performed as outlined in Section~\ref{sec:datatransmission}.

We first design a set $\mathcal{A}$ of $2^J$ preambles, each of length $n=MN$.
The $j$th active user picks an index $i_j \in [1,2^{J}]$ and transmits the preamble sequence $\mathbf{s}_{i_j}$ in the RACH slot. 
Let  $\mathbf{z}^{(\tau,\nu)}$ denote the preamble $\mathbf{z}$ shifted in time by $\tau$ and in frequency by $\nu$. 
We assume that the $j$th user's channel is a multipath channel with $L_j$ paths and hence,
the received signal can be expressed as 
\begin{equation}
    \mathbf{y} = \sum_{j=1}^K \sum_{l=1}^{L_j} h_{j,l} \mathbf{z}_{i_j}^{(\tau_{j,l},\nu_{j,l})} + \mathbf{n},
\end{equation}
where $\tau_{j,l}$, $\nu_{j,l}$, and $h_{j,l}$ denote the delay, Doppler, and attenuation associated with the $l$th path for the $j$th user.
The preamble detection problem can be cast as a sparse recovery problem.
The numerical results reported in this section are performed for the Veh-A channel specified in Section \ref{sec:datatransmission}.

We first construct an observation (or, sensing) matrix $\mathbf{A}$ as follows. A delay set $\mathcal{T}$ is defined as
\begin{align}
    \mathcal{T} = \left\{0, \frac{\tau_p}{M}, \frac{2\tau_p}{M}, \cdots, \left\lceil\frac{\tau_{\max}M}{\tau_p}\right\rceil\frac{\tau_p}{M}\right\}.
    \label{tau_set}
\end{align}
A Doppler set $\mathcal{D}$ is defined as
\begin{align}
    \mathcal{D} = \left\{-\left\lceil\frac{\nu_{\max}N}{\nu_p}\right\rceil\frac{\nu_p}{N}, \cdots,  0, \cdots,\left\lceil\frac{\nu_{\max}N}{\nu_p}\right\rceil\frac{\nu_p}{N}\right\}.
    \label{nu_set}
\end{align}
Observe that the sets $\mathcal{T}$ and $\mathcal{D}$ are the delay and Doppler values sampled on the OTFS grid. Another set $\mathcal{S}$ is defined as $\mathcal{S} = \mathcal{T} \times \mathcal{D}$, where $\times$ denotes the Cartesian product of two sets, i.e., the set $\mathcal{S}$ contains the combinations of delays and (discretized) Doppler spreads that each path could have. For every Zadoff-Chu sequence $z_j$ and for each $(\tau_i, \nu_i) \in \mathcal{S}$ the $(j\vert\mathcal{S}\vert+i)^{\text{th}}$ column of $\mathbf{A}$ is
\begin{align}
    \mathbf{A}_{j\vert\mathcal{S}\vert+i} = h_{\mathrm{eff}, i} *_{\sigma} \mathbf{z}_j,
\end{align}
where $h_{\mathrm{eff}, i}$ is computed using \eqref{h_eff} for $(\tau_i, \nu_i) \in \mathcal{S}$. In the matrix $\mathbf{A}$ the first $\vert S\vert$ columns correspond to the first Zadoff-Chu sequence $\mathbf{z}_1$, the next $\vert S\vert$ columns to $\mathbf{z}_2$ and so on.
The activity vector $\mathbf{x}$ is composed of $2^J$ sections each of length $MN$ where the $j$th section corresponds to the $j$th user. 
If the $j$th user is active, $\mathbf{x}_j$ is a $L_j$ sparse vector with $h_{j,l}, l = 1,2,\ldots, L_j$ being the non-zero entries. 
It can be seen that $\mathbf{x}$ is a block-sparse vector.
The received signal can be then be described as 
\begin{equation}
\label{eqn:MAassparserecovery}
    \mathbf{y} = \mathbf{A} \mathbf{x} +\mathbf{n}.
\end{equation}

\begin{figure}
    \centering
    \includegraphics[width=\linewidth]{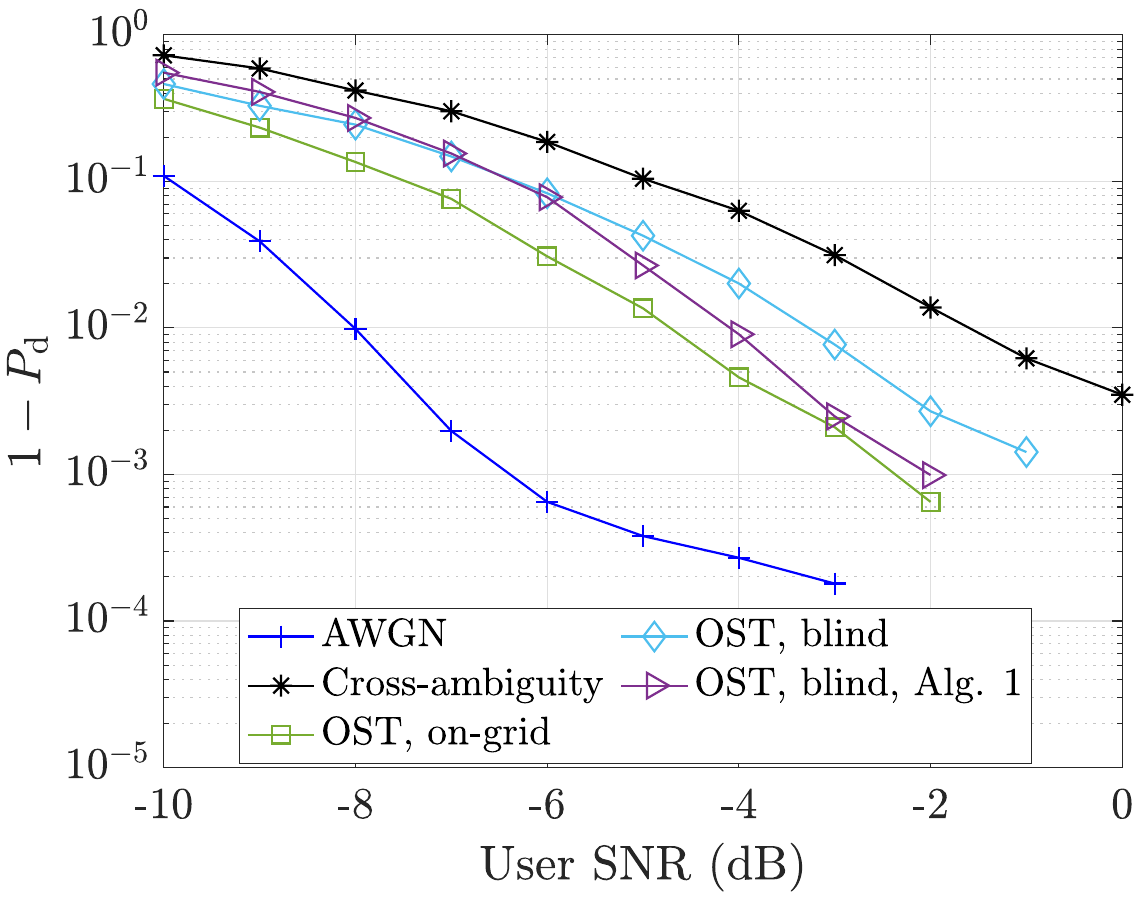}
    \caption{Probability of missed detection ($1-P_d$) as a function of user SNR. $M=31, N=37$ for a Zak-OTFS grid with Doppler period $\nu_p = 30$ KHz. Veh-A channel, RRC pulse shaping filter ($\beta_\tau = \beta_\nu = 0.6$), $\nu_{\max} = 815$ Hz. The performance of One Step Thresholding (OST) demonstrates the value of using a sparse prior for multiple preamble detection. The performance of OST when the receiver has no knowledge of the physical channel (OST, blind) is only slightly worse than that of OST when the path delays are known to the receiver (OST, on-grid).}
    \label{fig:prob_misdet}
\end{figure}

The one-step thresholding (OST) algorithm, presented in Algorithm \ref{alg:ost}, is used to solve the sparse recovery problem. 
The inputs to the algorithm are the observation matrix, the received vector, number of simultaneous users, set of delay and Doppler values, and a threshold. In the first step, a signal proxy is obtained by multiplying the hermitian of the observations matrix and the received vector. 
An index set $\mathcal{J}$ is generated to store the indices of all the ZC sequences. For each ZC sequence indexed by $j \in \mathcal{J}$, the indices of all the translates of the ZC sequence in $\mathbf{A}$ are stored in the set $\mathcal{J}_j$. A constituent sum $\vert\mathbf{f}_i\vert^2$ is obtained by summing over all $i \in \mathcal{J}_j$ for every ZC sequence. The indices corresponding to the constituent sums that are greater than the threshold $\lambda$ are declared as the indices of the detected preambles.
The probability of missed detection performance is presented in Fig. \ref{fig:prob_misdet} for various schemes. The AWGN and cross-ambiguity plots correspond to computing the cross-ambiguity between $\mathbf{y}$ and each ZC sequence (using \eqref{theorem3_eq1}) and picking the largest $K$ values for detecting preambles. OST, on-grid assumes perfect knowledge of delays and OST, blind computes OST without combining the energies of the paths (step 4 in Algorithm \ref{alg:ost}). The OST, blind, Alg. 1 is the plot obtained using the Algorithm \ref{alg:ost}. The performance of Algorithm \ref{alg:ost} is very close to that when delays are assumed to be known. Note that, even though all the users are assumed to have the same delays (see Table \ref{tab:veh_a}), the users are separated in Doppler. This separation along Doppler 
in conjunction with good correlation properties of the preambles allows the OST algorithm to achieve good performance.
Multi-user detection theory suggests that all the users having the same delays may be a worst-case scenario, and the performance of the OST algorithm will likely improve when the user delays are different. This remains to be studied.

It is natural to ask how we might use information about the physical channel gathered by the OST algorithm to detect the data transmitted in the corresponding resource blocks. 
In this paper, we do not explicitly consider this question. 
It is natural to use the data transmission scheme described in \ref{sec:datatransmission} and perform single-user detection. 
The result is a natural extension of the 2-step RACH \cite{peralta2021two} for doubly-dispersive channels and this provides a proof of concept that 
the 2-step RACH procedure described in the 5GNR standard can be operationalized for doubly-dispersive channels 
through the use of delay-Doppler signal processing.

Looking forward, we know from recent results in \cite{polyanskiy2017perspective,liva2024unsourced,agostini2024evolution}
that substantial improvements in the 2-step RACH procedure can be 
obtained by using unsourced random access instead of single-user decoding based slotted ALOHA for the Gaussian random access channel.
Indeed, a series of recent works \cite{liva2011graph,vem2019user,pradhan2022sparse,amalladinne2020coded,fengler2021sparcs,andreev2020polar,andreev2022coded,ahmadi2021random} has closed the gap between the performance of naive slotted ALOHA and the random coding benchmark of Polyanskiy \cite{polyanskiy2017perspective}.
Similar results are available somewhat scantily for the random access channel with quasi-static fading
\cite{kowshik2020energy,AlokSaif,fengler2021non,nassaji2021unsourced}.

The ability to detect multiple preambles, such as what is demonstrated in this paper, forms a key ingredient of unsourced random access schemes in \cite{vem2019user} and \cite{amalladinne2020coded}.
The ability to perform turbo detection as demonstrated in Section~\ref{sec:datatransmission} can be leveraged to design sparse IDMA-type random access schemes \cite{pradhan2022sparse}.
These approaches provide a natural path for designing sophisticated unsourced random access schemes for the data transmission part based on the delay-Doppler signal processing described in this paper.
This is a promising direction for bridging the gap between the performance of single-user based slotted ALOHA and the random coding bound, for doubly-dispersive channels.

\section{Conclusions}
\label{sec:conclusions}
We have provided a proof of concept that the 2-step RACH procedure described in the 5GNR standard can be realized for doubly-dispersive channels through the use of delay-Doppler signal processing. We have described a 2-step procedure that uses Zadoff-Chu sequences as preambles that point to radio resources subsequently used to upload data. We have demonstrated how to detect multiple preambles in the presence of mobility and delay spread using a receiver with no knowledge of the channel other than the worst case delay and Doppler spreads. We have described how to acquire the user I/O relation to support uplink data transmission in the presence of mobility and delay spread. Sensing the I/O relation and data transmission take place in the same subframe in order to maximize effective throughput. We avoid dividing radio resources between sensing and communication by designing pilot waveforms and carrier waveforms that are mutually unbiased so that one signal looks like noise to the other. These demonstrations make use of mathematical properties of Zadoff-Chu sequences derived in this paper. Prior work has demonstrated the potential of random coding given the simplifying assumption of a Gaussian random access channel. We look forward to future work demonstrating the same benefits in the presence of mobility and delay spread.

 





\end{document}